\documentclass[a4paper]{amsart}

\usepackage[dvips]{graphicx}
\usepackage{float}
\usepackage{latexsym}
\usepackage{amsmath,mathrsfs,bm,amssymb,color}
\usepackage{bbm}
\usepackage{hyperref}
\usepackage{amsfonts}
\usepackage{stmaryrd}
\definecolor{Blue}{rgb}{0.,0.,1.}
\definecolor{Red}{rgb}{1.,0.,0.}

\usepackage[T1]{fontenc}

\def\init{\setcounter{equation}{0}}

\newtheorem{theoreme}{Theorem }[section]
\newtheorem{proposition}[theoreme]{Proposition}
\newtheorem{lemma}[theoreme]{Lemma}
\newtheorem{definition}[theoreme]{Definition}
\newtheorem{corollary}[theoreme]{Corollary}
\newtheorem{remark}[theoreme]{Remark}

\newcommand\bbone{\ensuremath{\mathbbm{1}}}

\def\rr{\mathbb{R}}
\def\cc{\mathbb{C}}
\def\nn{\mathbb{N}}

\def\one{\bbone}

\newcounter{smallarabics}
\newenvironment{arabicenumerate}
{\begin{list}{{\normalfont\textrm{(\arabic{smallarabics})}}}
 {\usecounter{smallarabics}\setlength{\itemindent}{0cm}
 \setlength{\leftmargin}{5ex}\setlength{\labelwidth}{4ex}
 \setlength{\topsep}{0.75\parsep}\setlength{\partopsep}{0ex}
 \setlength{\itemsep}{0ex}}}
{\end{list}}

\newcounter{smallroman}

\newcommand{\ben}{\begin{arabicenumerate}}
\newcommand{\een}{\end{arabicenumerate}}

\def\e{{\rm e}}

\def\i{{\rm i}}
\def\d{{\rm d}}
\def\12{\frac{1}{2}}

\def\proof{\noindent{\bf Proof. }}
\def\slim{\hbox{\rm s-}\lim}

\def\coinf{C_{0}^{\infty}}

\def\qed{$\Box$}

\def\supp{{\rm supp\,}}
\def\cH{{\mathcal H}}

\def\cK{{\mathcal K}}

\def\K{{\mathcal K}}

\def\ch{{\mathfrak h}}

\def\p{\partial}
\def\s{{\rm s}}

\def\x{\langle x\rangle}

\def\pfi2{P(\varphi)_{2}}

\newcommand{\beq}{\begin{equation}}
\newcommand{\eeq}{\end{equation}}
\newcommand{\bet}{\begin{theoreme}}
\newcommand{\eet}{\end{theoreme}}
\newcommand{\bel}{\begin{lemma}}
\newcommand{\eel}{\end{lemma}}
\newcommand{\bep}{\begin{proposition}}
\newcommand{\eep}{\end{proposition}}
\newcommand{\bear}[1]{\begin{array}{#1}}
\newcommand{\ear}{\end{array}}

\newcommand{\ms}[1]{\mathscr #1}
\newcommand{\bd}[1]{\begin{definition}\label{#1}}
\newcommand{\ed}{\end{definition}}
\newcommand{\La}{\Lambda}
\newcommand{\bl}[1]{\begin{lemma}\label{#1}}
\newcommand{\el}{\end{lemma}}
\newcommand{\bc}[1]{\begin{corollary}\label{#1}}
\newcommand{\ec}{\end{corollary}}
\newcommand{\bt}[1]{\begin{theorem}\label{#1}}
\newcommand{\et}{\end{theorem}}
\newcommand{\bp}[1]{\begin{proposition}\label{#1}}
\newcommand{\ep}{\end{proposition}}
\newcommand{\br}[1]{\begin{remark}\label{#1}}
\newcommand{\er}{\end{remark}}
\newcommand{\EE}{\mathbb E}

\newcommand{\ix}[1]
{\int \mup {\EE}^\rx
\left[
#1
\right]}

\newcommand{\q}{{\rm q}}

\newcommand{\WW}{{\mathcal W}}

\newcommand{\ov}[1]{\overline{#1}}
\newcommand{\mup}{d\mu_{\rm p}(\rx)}

\newcommand{\WTT}{
\int_{-T}^T ds
\int_{-T}^T dt
W(X_s,X_t,|s-t|)
}

\newcommand{\WT}{
\int_{-T}^0 ds
\int_0^T dt W(X_s,X_t,|s-t|)
}

\newcommand{\WTZ}{
C_1 \int_{-T}^0 ds
\int_0^T dt W_\infty(X_s,X_t,C_2|s-t|)
}

\newcommand{\wtt}{
\int_{-T}^T ds
\int_{-T}^T dt W}

\newcommand{\eq}[1]{\begin{equation}\label{#1}}
\newcommand{\en}{\end{equation}}
\newcommand{\eqn}{\begin{eqnarray*}}
\newcommand{\enn}{\end{eqnarray*}}
\newcommand{\eqnn}{\begin{eqnarray}}
\newcommand{\ennn}{\end{eqnarray}}
\newcommand{\BR}{{{\mathbb R}^3 }}
\newcommand{\bi}{\begin{description}}
\newcommand{\ei}{\end{description} }

\newcommand{\kak}[1]{(\ref{#1})}

\newcommand{\HP}{{\hhh_{\rm p}}}

\newcommand{\LR}{{L^2(\BR)}}

\newcommand{\hz}{K_0}
\newcommand{\lp}{L}

\newcommand{\Ew}[1]
{\EE_{\WW}\lkkk#1\rkkk}

\newcommand{\wY}{\widetilde Y}

\newcommand{\is}{\inf\sigma}
\newcommand{\f}{^{-1}}
\newcommand{\lk}{\left(}
\newcommand{\rk}{\right)}
\newcommand{\lkk}{\left\{}
\newcommand{\rkk}{\right\}}
\newcommand{\lkkk}{\left[}
\newcommand{\rkkk}{\right]}

\newcommand{\ab}[1]{\langle#1
\rangle}

\renewcommand{\d}{\displaystyle}

\newcommand{\hf}{H_{\rm f}}

\newcommand{\hhh}{{\mathcal H}}

\newcommand{\grp}{\psi_{\rm p}}

\newcommand{\half}{\frac{1}{2}}
\newcommand{\han}{{1/2}}

\newcommand{\pro}[1]{(#1_t)_{t\geq0}}

\newcommand{\non}{\nonumber}
\newcommand{\FFF}{{\mathcal F}}

\newcommand{\VV}{V}
\newcommand{\YYY}{\mX_+}
\newcommand{\mU}{\ms U}
\newcommand{\mB}{B}
\newcommand{\mX}{\ms X}
\newcommand{\rP}{{\rm P}}
\newcommand{\rQ}{{\rm Q}}

\newcommand{\ixp}[1]
{\int \mup {\EE}_{{\rP^{\rx}}}
\left[#1\right]}

\begin{document}
\def\triple{\interleave}
\def\Gh{\Gamma(\ch)}
\def\Dom{{\rm Dom}}
\def\y{\langle y\rangle}
\def\rx{{\rm x}}
\def\ry{{\rm y}}
\def\cX{{\mathcal X}}

\title{Absence of ground state \\ for  the Nelson model on static space-times}
\author{C. G\'erard}
\address{D\'epartement de Math\'ematiques, Universit\'e de Paris XI, 91405 Orsay Cedex France}
\email{christian.gerard@math.u-psud.fr}
\author{F. Hiroshima}
\address{Department of Mathematics, University of Kyushu, 6-10-1, Hakozaki, Fukuoka, 812-8581, Japan}
\email{hiroshima <hiroshima@math.kyushu-u.ac.jp>}
\author{A. Panati}
\address{PHYMAT, Universit\'e Toulon-Var 83957 La Garde Cedex France}
\email{annalisa.panati@univ-tln.fr}
\author{A. Suzuki}
\address{Department of Mathematics, Faculty of Engineering, Shinshu University, 4-17-1 Wakasato, Nagano 380-8553, Japan}
\email{sakito@math.kyushu-u.ac.jp}
\date{November 2010}

\begin{abstract}
We consider  the Nelson model on some static space-times
and investigate the problem of absence of a ground state.
Nelson models with variable
coefficients arise when one replaces in the usual Nelson model
the flat Minkowski metric by a  static metric, allowing also the
boson mass to depend on position.   We investigate the
absence of a ground state of the Hamiltonian in the presence of the
infrared problem, i.e. assuming that the boson  mass  $m(x)$ tends to $0$ at spatial
infinity. Using path space techniques, we show that if $m(x)\leq C |x|^{-\mu}$ at infinity for some $C>0$  and $\mu>1$
then the Nelson Hamiltonian has no  ground state.\end{abstract}
\maketitle


\section{Introduction}
In this paper we continue the study of the so-called {\em Nelson model with  variable coefficients}  began in \cite{ghps1, ghps2}.
The Nelson model with variable coefficients
describes a system of quantum particles linearly coupled
to a scalar quantum field with an ultraviolet cutoff. Typically the scalar field is the Klein-Gordon field on a static Lorentzian manifold, (see \cite{ghps2}).

In this respect the Nelson model with variable coefficients  is an extension of the standard Nelson model introduced by \cite{ne}
to the case when the Minkowskian space-time is replaced by a static Lorentzian manifold.

The Hamiltonian of the Nelson model
with variable coefficients is defined
as a selfadjoint operator on
$L^2(\rr^3,d\rx)\otimes
\Gamma_{\s}(L^2(\rr^3,dx))$,
formally given by
\begin{align}
H&=
-\half \sum_{1\leq j,k\leq 3}
\p_{\rx_j}A^{jk}(\rx)\p_{\rx_k}+\VV(\rx)
\label{g123}\\
&+\half \int
\lk
\pi(x)^2+\varphi(x) \omega^2(x,D_{x}) \varphi(x)
\rk
dx \nonumber \\
&
+\frac{q}{\sqrt2}
\int \omega^{-\han}(x,D_{x})
 \rho(x-\rx)
\varphi(x) dx,
\nonumber
\end{align}
where $\varphi(x)$ is the time-zero scalar field,
$\pi(x)$ its conjugate momentum,
$q\in \rr $ a coupling constant,
$\rho $ a non-negative cutoff function,
and $\omega(x, D_{x})=h^{\12}$
with
\eq{g234}
h=-c(x)^{-1}
\lk \sum_{1\leq j,k\leq 3}\p_{x_j}a^{jk}(x)\p_{x_k}\rk c(x)^{-1}+m^2(x).
\en
Here
$m^2(x) $ describes a variable mass.
The assumptions  on $a^{jk}$, $A^{jk}$ and $c$ will be  given later in Section  \ref{arsu}. We
refer to  \cite{ghps2} for
the derivation
of \kak{g123}  starting from the  Lagrangian of a  Klein-Gordon field on a static space-time linearly coupled to a non-relativistic  particle.

The standard Nelson model is defined by taking $\omega(x, D_{x})= \omega(D_{x})$ for $ \omega(k)= (k^{2}+ m^{2})^{\12}$ with a constant $m\geq 0$,
and $A^{jk}=\delta_{jk}$.
Then $m>0$ (resp.  $m=0 $) corresponds to the massive (resp.  massless) case. The model is called {\em infrared singular} (resp.  {\em regular}) if
\[
\int_{\rr^{3}} \frac{|\hat\rho (k)|^2}{\omega(k)^3}dk=\infty \hbox{ (resp. }<\infty\hbox {)},
\]
in particular the massive case is always infrared regular. In this paper  we will assume that   $\rho \geq0$ and $\int_{\rr^{3}}\rho(x)d x=1$,
which in the standard Nelson model leads to an
 infrared singular interaction
(see Remark \ref{irs}).
In the infrared regular case, it is now well known that the standard Nelson Hamiltonian has a unique
 ground state,
see \cite{bfs2, dg2, ggm, ger00, spo98} and \cite{hhs, hs08, pan,sas} for more general results.
The ground state properties
are discussed in \cite{bhlms} using path space techniques.
It is also known  that  in the infrared singular case the  standard Nelson Hamiltonian  has no ground state.
See \cite{ahh99, hi06, lms,dg1}.

In \cite{ghps2} the existence of ground states of $H$
is shown when
\eq{g4}
m(x)\geq C \ab{x}\f, \ C>0,
\en
where $\ab{x}=(1+|x|^2)^\han$.    In this paper we will consider the case
\begin{equation}\label{carla}
m(x)\leq C \ab{x}^{-\mu}, \ \mu>1.
\end{equation}
In \cite{ghps1},
the absence of ground state
of the Nelson model \kak{g123}
is proven if (\ref{carla}) holds for
 $\mu>3/2$, for a sufficiently small coupling constant,
and
$A^{jk}(\rx)=\delta_{jk}=a^{jk}(x)$.
In the present paper we drastically extend \cite{ghps1}.
In fact we  show that if (\ref{carla}) holds for some $\mu>1$ then $H$ has no ground state.
Therefore  combining the results of \cite{ghps2} with those of the present paper gives an essentially complete
discussion of the problem of existence of a ground state for the Nelson model with variable coefficients.

In \cite{dg1} the absence of ground states for an abstract class of models  including the standard Nelson model is shown by making use of the so-called \emph{pull-through formula}.  This method does not seem to be applicable in our situation.
%
%
Instead we use
 the method developed in \cite{lms} based on path space arguments.
 We now briefly explain this approach.

\medskip

{\bf Path space representation of the Nelson model.}

\medskip

One can write the physical Hilbert space $L^{2}(\rr^{3})\otimes \Gamma_{\rm s}(L^{2}(\rr^{3}))$ as  $L^{2}(M,dm)$ for some probability space $(M,m)$  in such a way that the interaction term $\varphi_{\rho}(\rx)$ becomes a multiplication operator on $M$ and the semi-group $\e^{- tH}$ is positivity improving.  Moreover the expectation values $(F|\e^{- tH}G)$ can be written using an appropriate path space measure and a Feynman-Kac formula, and the ground state of the free Hamiltonian $H_{0}$, (i.e. $H$ with $q=0$),  is mapped to 
the constant function $\one$.

The  probability space $(M,m)$ and the path space measure are obtained by tensoring the corresponding objects for the particle and field Hamiltonians. For the particle Hamiltonian $K$ we use  the fact that $K$ has a strictly  positive ground state $\varphi_{\rm p}$. We then apply
 the so called {\em ground state transform} by unitarily identifying $L^{2}(\rr^{3}, d \rx)$ with $L^{2}(\rr^{3}, \grp (\rx) d\rx)$, obtaining a new particle Hamiltonian $L$. One can   then construct a diffusion process associated to  the semi-group $\e^{- tL}$.

 For the field Hamiltonian we use the well-known Gaussian process. The path space representation for the Nelson model is then obtained from a Feynman-Kac-Nelson  formula.

\medskip

{\bf Absence of ground state.}

\medskip

After mapping everything to $L^{2}(Q, \d \mu)$, an easy argument based on Perron-Frob\-enius shows that  $H$ has no ground state iff
\[
\gamma(T):=\frac{(\one| \e^{-TH}\one)^2}{(\one| \e^{-2TH}\one)}
\]
 tends to $0$ when $T\to +\infty$. Using the Feynman-Kac formula the expectation value $(\one|\e^{- TH}\one)$ can be explicitly expressed in terms of the
\emph{pair potential} $W$ given by
$$W (\rx,\ry,|t|)=
(\rho (\cdot-\rx)|
\frac{\e^{-|t|\omega}}{2\omega} \rho (\cdot-\ry)).$$
The key ingredient to estimate $W$ are
Gaussian bounds such as
 $$C_1\e^{tC_2\Delta}(x,y)\leq \e^{-t\omega^2}(x,y)\leq C_3\e^{tC_4\Delta}(x,y).$$
 By modifying the method used in
 \cite{lms,var} and using the super-exponential decay of $\grp$,
 we can finally show that
$\gamma(T)\to0$ as $T\to \infty$ and we conclude that $H$ has no ground state.

\medskip

{\bf Organization.}

This paper is organized as follows. In Section 2 we define the Nelson Hamiltonian
with variable coefficients.
In Section 3 we consider the semi-groups $\e^{- tK}$ and $\e^{- tL}$ associated to the two versions of the particle Hamiltonian.  We prove the Feynman-Kac formula and various Gaussian  bounds on
$\e^{- tK}$ and $\e^{- th}$.  We also
construct the diffusion process associated with $\e^{-t\lp}$.
In Section 4 the functional integral representation of $\e^{-tH}$ is given.
In Section 5 we prove the absence of ground state.
Finally Appendix \ref{a4} is devoted to the proof of  Proposition  \ref{main1} about the diffusion process associated with $\lp$.

\section{The Nelson model with variable coefficients}\label{arsu}
In this section we define the Nelson model with variable coefficients and state our main theorem.

\subsection{Notation}
We collect  here some notation used in this paper for reader's convenience.

{\bf Hilbert space and operators:}
The domain of a linear operator $A$ on Hilbert space $\cH$ will be denoted by $\Dom A$, and its spectrum by $\sigma(A)$. The set of
bounded operators from $\cH$ to $\cK$ is denoted by $B(\cH,\cK)$ and $B(\cH,\cH)$ by $B(\cH)$ for simplicity.
The scalar product on $\cH$ is denoted by $(u| v)$.
Let ${\mathcal X}$ be a real or complex Hilbert space. If $a$ is a
selfadjoint operator on $\cX$, we will write $a>0$ if $a\geq 0$
and ${\rm Ker}a=\{0\}$.
Note that if $a>0$ and
$s\in \rr$, $\|h\|_{s}= \|a^{-s}h\|_{\cX}$ is a norm on
$\Dom  a^{-s}$.
We denote then by $a^{s}{\mathcal X}$ the completion of $\Dom  a^{-s}$
for the norm $\|\ \|_{s}$.
The map $a^{s}$ extends as a
unitary operator from $a^{t}{\mathcal X}$ to $a^{s+t}{\mathcal X}$.
One example of this notation are the familiar  Sobolev spaces, where
$H^{s}(\rr^{d})$ is equal to $(-\Delta +1)^{-s/2}L^{2}(\rr^{d})$.
Finally if $B\in B(L^{2}(\rr^{3}))$, the distribution kernel of $B$ will be denoted by $B(x, y)$.

{\bf Bosonic Fock space:} If $\ch$ is a Hilbert space, the {\em bosonic Fock space} over $\ch$, denoted by
$\Gamma_{\rm s}(\ch)$, is
\[
\Gamma_{\rm s}(\ch):=\bigoplus_{n=0}^{\infty}\otimes_{\rm s}^{n}\ch.
\]
$\Omega=(1,0,0,\cdots)\in \Gamma_\s(\ch)$ is called the Fock vacuum.
We denote by $a^{*}(h)$ and $a(h)$
for $h\in \ch$
the {\em creation }
and {\em annihilation operators},
acting on $\Gamma_{\rm s}(\ch)$.
If $\cK$ is another Hilbert space and $v\in B(\cK, \cK\otimes\ch)$, then one defines the operators $a^{*}(v)$,
$a(v)$ and $\phi(v)$ as unbounded operators on $\cK\otimes \Gamma_{\rm s}(\ch)$ by
\[\begin{array}{l}
a^*(v)\Big|_{\cK\otimes(\otimes_\s^n\ch)}
:=\sqrt{n+1}\Big(\one_\cK\otimes
{\mathcal S}_{n+1}\Big)
\Big(v\otimes\one_{\bigotimes_\s^n\ch}\Big),\\[3mm]
a(v):=\big(a^*(v)\big)^*,\\[3mm]
\d \phi(v):=\frac{1}{\sqrt2}(a(v)+a^*(v).
\end{array}\]
Here ${\mathcal S}_{n+1}$ denotes the symmetrization.
If $T$ is a contraction on $\cH$,
then  $ \Gamma(T):\Gamma_\s(\ch)\to\Gamma_\s(\ch)$
is defined as
 \[
\begin{array}{l}
\Gamma(T)\Big|_{\bigotimes_\s^n\ch}:
=\underbrace{
T\otimes \cdots\otimes T}_{n}
,\quad n\geq1,\\
\Gamma(T)\Big|_{\bigotimes_\s^0\ch}:
=\one,\quad n=0.
\end{array}
\]
If $b$ is a selfadjoint operator on $\ch$, its
second quantization $ d\Gamma(b):\Gamma_\s(\ch)\to\Gamma_\s(\ch)$
is defined as
 \[
\begin{array}{l}
d\Gamma(b)\Big|_{\bigotimes_\s^n\ch}:
=\sum\limits_{j=1}^n\underbrace{\one\otimes\cdots\otimes\one}_{j-1}
\otimes b\otimes \underbrace{\one\otimes\cdots\otimes\one}_{n-j},\quad n\geq1,\\
d\Gamma(b)\Big|_{\bigotimes_\s^0\ch}:
=0,\quad n=0.
\end{array}
\]
Let $N=d\Gamma(\one)$.
The creation operator and the annihilation operators satisfy the estimates
\beq
\|a^{\sharp}(v)(N+1)^{-\12}\|\leq \|v\|,
\label{ju88}
\eeq
where $a^\sharp=a,a^\ast$ and $\|v\|$ is the norm of $v$ in $B(\K,\K\otimes\ch)$.

We denote by $x\in \rr^{3}$ (resp. $\rx\in \rr^{3}$) the boson (resp. particle) position.

\subsection{Particle Hamiltonian}\label{elec}
In this section we define
the particle Hamiltonian $K$ on $\LR$.
 We set
\[
K_{0}= -\12\sum_{1\leq j,k \leq 3}\p_{\rx_{j}}A^{jk}(\rx)\p_{\rx_{k}},
\]
acting on $\cK=L^{2}(\rr^{3}, d\rx)$. We assume
$$
\begin{array}
{ll}
(E1)& C_{0}\one \leq [A^{jk}(\rx)]\leq C_{1}\one,\ C_{0}>0,\\
(E2)& \nabla_{\rx}[A^{jk}(\rx)]\in L^{\infty}(\rr^{3}).
\end{array}
$$
In Subsection \ref{sec1.2} we will consider the drift vector:
\[
b(\rx)= (b_{1}(\rx), b_{2}(\rx), b_{3}(\rx)), \ b_{k}(\rx)= \12 \sum_{j=1}^{3}\p_{j}A^{jk}(\rx),
\]
and we will need the assumption:
\[
(E3)\ \ \nabla_{\rx} b_j(\rx)\in L^{\infty}(\rr^{3}).
\]
Under assumption (E1), $K_{0}$ is defined as the positive selfadjoint operator associated with the closed quadratic form:
\eq{q}
\q_{0}(f, f)=\12\int\sum_{1\leq j,k\leq3}
\overline{\p_{\rx_j}
f(\rx)}
A^{jk}(\rx)\p_{\rx_k}f(\rx) d\rx,
\en
with form domain $H^{1}(\rr^{3})$. Assuming also (E2), then by standard elliptic regularity, we know that
\[
 \ K_{0}f= -\12\sum_{1\leq j,k \leq 3}
 \p_{\rx_{j}}(A^{jk}(\rx)\p_{\rx_{k}}f)
\]
with
$\Dom K_{0}= H^{2}(\rr^{3})$.
We also introduce  an external potential
$V$.
We assume that
\[
(E4) \  V\in L^{1}_{\rm loc}(\rr^{3}), V\geq 0.
\]
The operator
\[
K:= K_{0}\,\, \dot +\,\, V
\]
is defined as the positive selfadjoint operator associated
with the closed quadratic form:
\[
\q(f, f)= \q_{0}(f,f)+ \int V(\rx)|f|^{2}(\rx) d\rx,
\]
with form domain
$H^{1}(\rr^{3})\cap \Dom V^{\12}$. If we assume the following {\em confining condition:}
\[
(E5) \ \ b_{0}\langle\rx\rangle^{2\delta}
\leq V(\rx),  \ b_{0}>0, \ \delta>0.
\]
then $K$ has compact resolvent.

\subsection{Boson one-particle energy}
Next we define boson one-particle Hamiltonian.
Let
\beq\label{irlop}
\begin{array}{rl}
h_{0}:=&
\displaystyle
- c(x)^{-1}\left(\sum_{1\leq j,k\leq d}\p_{j}a^{jk}(x)\p_{k}\right)c(x)^{-1},
\\[4mm]
 h:=& h_{0}+ m^{2}(x),
\end{array}
\eeq
where $a^{jk}$, $c$, $m$ are real functions and
\[
(B1) \ \begin{array}{l}
C_{0}\one\leq [a^{jk}(x)]\leq C_{1}\one, \ C_{0}\leq c(x)\leq C_{1}, \ C_{0}>0, \\[2mm]
\p_{x}^{\alpha}a^{jk}(x)\in O(\langle x\rangle^{-1}), \ |\alpha|\leq 1, \\[2mm]
\p_{x}^{\alpha}c(x)\in O(1), \
|\alpha|\leq 2,\\[2mm]
\p_{x}^{\alpha}m(x)\in O(1), \ |\alpha|\leq 1.
\end{array}
\]
We assume that the variable mass term $m(x)$ decays at infinity faster than $\x^{-1}$:
\[
(B2)\ m(x)\in O(\x^{-\mu}), \ \mu>1.
\]
Clearly $h$ is selfadjoint on $H^{2}(\rr^{3})$ and
$h\geq 0$. The {\em one-particle space} and
{\em one-particle energy} are
\beq\label{ash}
\ch:=L^2(\rr^3,dx),\quad
\omega:=h^{\12}.
\eeq
By \cite[Lemma 3.1]{ghps2} we know that
\[
{\rm Ker}\omega=\{0\}, \ {\rm inf}\sigma(\omega)=0.
\]
\subsection{Nelson Hamiltonians}
We fix a {\em charge density} $\rho: \rr^{3}\to \rr$ such that
\[
(B3)
\ \rho(x)\geq 0,\
\int \rho(x) dx=1,
\ |k|^{-\alpha}\hat{\rho}(k)
\in L^{2} (\rr^{3}, d k),\ \alpha=1,\12.
\]
where $\hat \rho$ denotes the Fourier transform of $\rho$,
and set
$\rho_{\rx}(x)= \rho(x-\rx)$.
We define the {\em UV cutoff fields} as
\beq\label{e0.1}
\varphi_{\rho}(\rx):=
\phi( \omega^{-\12} \rho_{\rx}).
\eeq
Note that setting
$
\varphi(\rx):= \phi(\omega^{-\12}\delta_{\rx})$,
one has $\varphi_{\rho}(\rx)= \int \varphi(\rx-\ry)\rho(\ry)d \ry$.
The {\em Nelson Hamiltonian} is
\beq\label{e2.1}
H:= K\otimes\one + \one \otimes {\rm d} \Gamma(\omega)+ q\varphi_{\rho}(\rx),
\eeq
acting on
\eq{z1}
\cH= \cK\otimes\Gamma_{\rm s}(\ch).
\en
The constant $q$ has the interpretation of the charge of the particle. We assume of course that
$q\neq 0$.
We also set
\[
H_{0}:=K\otimes\one + \one \otimes {\rm d}\Gamma(\omega),
\]
which is selfadjoint on $\Dom H_{0}= \Dom (K\otimes\one) \cap \Dom(\one\otimes {\rm d}\Gamma(\omega))$.
\begin{proposition}
 Assume hypotheses (E1), (E4), (B1), (B2), (B3). Then
$H$ is selfadjoint and bounded below on $\Dom H_{0}$. Moreover $H$ is essentially selfadjoint on any core of $H_{0}$.
\end{proposition}

\proof Since $-\Delta_{x}\leq C \omega^{2}$ it follows from the Kato-Heinz theorem that
\[
\sup_{\rx\in \rr^{3}}\|\omega^{- \alpha}\rho_{\rx}\|_{L^{2}}\leq C \||k|^{- \alpha}\hat{\rho}\|_{L^{2}}, \ \alpha= \12, 1.
\]
It follows e.g., from \cite[Section 4]{ggm} that $\varphi_{\rho}(\rx)$ is $H_{0}$ bounded with the infinitesimal bound, and the proposition follows from the Kato-Rellich theorem. \qed

\medskip

\begin{remark}
\label{yoko22}
 In the previous paper \cite{ghps1} we considered the case \\ $\omega= (- \Delta_{x}+ m^{2}(x))^{\12}$ but with
 $\varphi_\rho $
replaced by $\tilde \varphi_\rho $ given by
\eq{yoko23}
\tilde \varphi_\rho (\rx)=\phi(\omega^{-\half}\tilde{\rho}_\rx),\en
where
\eq{ck1}
\tilde{\rho}_\rx(\cdot)=(2\pi)^{-3/2}
\int \Psi(k,\cdot)\ov{\Psi(k,\rx)}
\hat \rho (k) dk,
\en
 and the generalized eigenfunctions $\Psi(k,x)$ are solutions to
the Chapman-Kolmogor\-ov equation: \eq{ck}
\Psi(k,x)=\e^{\i k\cdot x}-\frac{1}{4\pi}
\int_\BR
\frac{\e^{\i|k||x-y|}m^2(y) \Psi(k,y)}{|x-y|}dy.
\en
Note that if $\rho$ is radial e.g., $\rho(x)= \rho(|x|)$ then $\tilde{\varphi}_{\rho}(x)= \phi(\omega^{-\12}\hat{\rho}(\omega)\delta_{\rx})$.
If $m(x)\equiv 0$ then
$\tilde \varphi_\rho (\rx)=\varphi_\rho (\rx)$.
In the general
case, $\omega=h^\half$,
the natural definition
\kak{e0.1} is much more convenient than \kak{yoko23}.
In particular we do not need to
consider generalized eigenfunctions for $h$ defined in (\ref{irlop}).
\end{remark}
\subsection{Absence of ground state for Nelson Hamiltonians}
The main theorem in this paper is as follows:
\bet\label{hiroshima4}
Assume hypotheses (E1), (E2), (E3), (E5), (B1), (B2) and (B3). Then $H$ has no ground state.
\eet

\begin{remark}
\label{irs}
Since $\hat \rho (0)=1$, we see that
\eq{ir}\int_\BR\frac{|\hat\rho (k)|^2}{|k|^3}dk
=\infty.
\en
As is well known if $\omega= (-\Delta_{x})^{\12}$,
\kak{ir} is called the {\em infrared singular condition}.
In this case Theorem \ref{hiroshima4} is well known, see e.g., \cite{dg1}.
\end{remark}\begin{remark}
 In \cite{ghps2} we show that if instead of (B2) we assume that $m(x)\geq C \x^{-1}$ then $H$ as a (unique) ground state. Therefore Theorem \ref{hiroshima4} is sharp with respect to the decay rate of the mass at infinity.
\end{remark}

\section{Feynman-Kac formula for the particle Hamiltonian}

In this section we prove some Gaussian bounds on the heat kernels $\e^{- tK_{0}}$, $\e^{- th_{0}}$ and
$\e^{- th}$. The bounds for
$\e^{- tK_{0}}$ and
$\e^{- th_{0}}$ are well known in various contexts. In our situation they are due to 
\cite[Theorems 3.4 and 3.6]{PE}.
Note that by identifying $x$ and $\rx$
and setting $c(x)\equiv 1$,
 $K_{0}$ is a particular case of $h_{0}$.

\subsection{Gaussian upper and lower bounds on heat kernels}
The bounds for
$\e^{- th}$ were proved previously by \cite{se} for operators in divergence form and by \cite{Zh}
for Laplace-Beltrami operators.

\begin{proposition} \cite[Theorems 3.4 and  3.6]{PE}\label{irot}
 Assume (B1), or (E1), (E2). Then there exist constants $C_{i}, \ c_{i}>0$ such that
 \begin{equation}
\label{bibis}
C_{1}
\e^{ c_{1}t\Delta}(x, y)\leq
\e^{- th_{0}}(x, y)\leq C_{2}
\e^{ c_{2}t\Delta}(x, y), \ \forall \ t>0, \ x, y\in \rr^{3},
\end{equation}
 \begin{equation}
\label{bib}
C_{1}\e^{ c_{1}t\Delta}(\rx, \ry)\leq \e^{- tK_{0}}(\rx, \ry)\leq C_{2}\e^{ c_{2}t\Delta}(\rx, \ry), \ \forall \ t>0, \ \rx, \ry\in \rr^{3}.
\end{equation}
\end{proposition}
\begin{proposition}\label{lowbound}
 Assume (B1) and  (B2).
Then there exist constants $C_{i}, c_{i}>0$ such that
\[
C_{1}\e^{ c_{1}t\Delta}(x, y)\leq \e^{- th}(x, y)\leq C_{2}\e^{ c_{2}t\Delta}(x, y), \ \forall \ t>0, \ x, y\in \rr^{d}.
\]
\end{proposition}
\begin{remark}\label{ilt}
Conjugating by the unitary
$$
U:
L^{2}(\rr^{d}, d x)\ni u\mapsto c(x)^{-1}u\in L^{2}(\rr^{d}, c^{2}(x)d x),
$$
we obtain
\[
\begin{array}{rl}
\tilde{h}_{0}:=&Uh_{0}U^{-1}= -c(x)^{-2}\sum_{1\leq j,k\leq d} \p_{j}a^{jk}(x)\p_{k}, \\[2mm]
 \tilde{h}:=&UhU^{-1}= \tilde{h}_{0}+ m^{2}(x),
\end{array}
\]
which are selfadjoint with domain $H^{2}(\rr^{d})$.
 Let $\e^{-t\tilde{h}}(x, y)$ for $t>0$ the integral kernel of $\e^{-t\tilde{h}}$ i.e. such that
 \[
\e^{-t\tilde{h}}u(x)=\int_{\rr^{d}}\e^{-t\tilde{h}}(x, y)u(y)c^{2}(y)d y, \ t>0.
\]
Then since $\e^{-th}(x, y)= c(x)\e^{-t\tilde{h}}(x, y)c(y)$,  the bounds in Proposition  \ref{irot} also hold for $\tilde{h}_{0}$ and
it suffices to prove Proposition \ref{lowbound} for $\e^{-t\tilde{h}}$.
\end{remark}
By the above remark, we will consider the operators $\tilde{h}_{0}$  and  $\tilde{h}$.
We note that they are associated with the closed quadratic forms:
\beq\label{harie}
\begin{array}{ll}
\d \tilde{h}_{0}(f,f)=\int_{\rr^{d}}
\sum_{1\leq j,k\leq3}
\overline{\p_{j}{f}(x)}a^{jk}(x)\p_{k}f(x)\ d x, \\
\d \tilde{h}(f,f)=
\tilde{h}_{0}(f,f)+
\int_{\rr^{d}}
|f|^{2}(x)m^{2}(x)c^2(x) \ d x,
\end{array}
\eeq
with domain $H^{1}(\rr^{d})$.
We will use the following well known convexity result. For completeness we sketch its proof below.
\begin{lemma}\label{convex}
Assume that $w\in L^{\infty}(\rr^{3})$ is a real potential. Then
\[
\rr\ni \lambda\mapsto \e^{-t(\tilde{h}_{0}+ \lambda w)}(x, y)
\]
is logarithmically convex for all $t>0$ and a.e. $x, y\in \rr^{3}$.
\end{lemma}
\proof
Note that if $F_{1}$ and $F_{2}$ are log-convex, then $F_{1}F_{2}$ and $CF_{1}$ are log-convex. Moreover (see \cite[Theorem 13.1]{sim04}) if for all $y\in \rr^d$ the function $\rr\ni \lambda\mapsto F(\lambda, y)$ is log-convex, so is $\lambda\mapsto \int_{\rr^{d}} F(\lambda, y)d y$.
To prove the claim we use the Trotter product formula. We can set $t=1$.
\[
\e^{-(\tilde{h}_{0}+ \lambda w)}(x, y)=\lim_{n\to \infty} (\e^{- \tilde{h}_{0}/n}\e^{- \lambda w/n})^{n}(x,y),\hbox{ a.e.} \ x, y.
\]
Let $A_{\lambda}( x, y)$, $B_{\lambda}( x, y)$ the kernels of two operators $A_\lambda, B_{\lambda}$ assumed to be log-convex in $\lambda$ for a.e. $x, y$. Then by the above remarks
the kernel of $A_{\lambda}B_{\lambda}$
\[
(A_{\lambda}B_{\lambda})(x, y)=\int_{\rr^{d}} A_{\lambda}( x, y')B_{\lambda}(y', y)d y'
\]
is also log-convex in $\lambda$. The kernel of $\e^{-\tilde{h}_{0}/n}
\e^{-\lambda w/n}$ equals to
$\e^{- \tilde{h}_{0}/n}(x, y)
\e^{- \lambda w(y)/n}$ is log-convex in $\lambda$. Applying the above remark and the Trotter formula we obtain our claim. \qed

\medskip

{\bf Proof of Proposition \ref{lowbound}}.
We use the unitary transformation as in Remark \ref{ilt}.
We know from Proposition \ref{irot} that
\begin{equation}
\label{bib2}
C_{1}\e^{ c_{1}t\Delta}(x, y)\leq \e^{- t\tilde{h}_{0}}(x, y)\leq C_{2}\e^{ c_{2}t\Delta}(x, y), \ \forall \ t>0, \ x, y\in \rr^{d}.
\end{equation}
Since $m^{2}(x)\geq 0$, the upper bound in the proposition follows from the Trotter-Kato formula. Let us now prove the lower bound, following the arguments in \cite[Theorem 6.1]{se}.
Since $m^{2}(x)c^2(x)\in O(\langle x\rangle^{-2- \epsilon})$ it follows from Hardy's inequality and (\ref{harie}) that
\[
\tilde{h}_{0}\geq c_{0}m^{2},
\hbox{ for some }c_{0}>0.
\]
Set now $w(x)= -c_{0}m^{2}(x)/4$. Since $\tilde{h}_{0}+ 2w\geq \12 \tilde{h}_{0}$, we deduce from \cite[Theorem 2.4.2]{Dav} that
\[
\|\e^{-t (\tilde{h}_{0}+ 2 w)}\|_{L^{2}\to L^{\infty}}\leq C t^{-d/4}.
\]
By duality this implies that
\[
\|\e^{-t (\tilde{h}_{0}+ 2 w)}\|_{L^{1}\to L^{2}}\leq C t^{-d/4},
\]
and hence
\[
\|\e^{-t (\tilde{h}_{0}+ 2 w)}\|_{L^{1}\to L^{\infty}}\leq \|\e^{-t (\tilde{h}_{0}+ 2 w)/2}\|_{L^{2}\to L^{\infty}}\|\e^{-t (\tilde{h}_{0}+ 2 w)/2}\|_{L^{1}\to L^{2}}\leq Ct^{-d/2}.
\]
By \cite[Lemma 2.1.2]{Dav} we obtain
\[
\e^{- t(\tilde{h}_{0}+ 2w)}(x,y)\leq C t^{-d/2}.
\]
Applying then Lemma \ref{convex} this yields
\beq\label{bab}
\e^{- t(\tilde{h}_{0}+w)}(x, y)\leq t^{-d/4}\e^{- t h_{0}}(x, y)^{\12}, \hbox{ a.e. }x, y\in \rr^{d}.
\eeq
Applying once more the log-convexity, we get that
\[
\e^{- t\tilde{h}_{0}}(x, y)\leq \e^{- t(\tilde{h}_{0}+ v)}(x, y)^{s}\e^{- t(\tilde{h}_{0}+ w)}(x, y)^{1-s}, \hbox{ for }s=c_{0}/(4+ c_{0}),
\]
and hence using (\ref{bab}):
\[
\e^{- t\tilde{h}_{0}}(x, y)^{(1+s)/2}t^{(1-s)d/4}\leq \e^{-t(\tilde{h}_{0}+m^{2}c^2)}(x, y)^{s},
\]
which implies the lemma using the lower bound in (\ref{bib2}). \qed

\subsection{Stochastic differential equation}\label{sec1.2}
Recall that we introduced the drift vector $b(\rx)$ in
Subsection \ref{elec}. We also define the diffusion matrix:
\[
\sigma(\rx):= [A^{jk}]^{\12}(\rx),
\]
and note that it follows from (E1), (E2), (E3) that $b(\rx), \ \sigma(\rx)$ are
uniformly Lipschitz on $\rr^{3}$.
We consider the stochastic differential equation:
\begin{equation}\label{n1}
\left\{
\begin{array}{lcl}
dX_t^\rx&=&b(X_t^{\rx}) dt+\sigma(X_t^{\rx}) dB_t,\ t\geq 0,\\
 X_0^{\rx}&=&\rx,
\end{array}\right.
\end{equation}
on the probability space
$(\YYY , \mB(\YYY), \WW)$,
where
$\YYY =C([0,\infty);\BR)$,
$\mB(\YYY)$ is the $\sigma$-field generated by cylinder sets and
$\WW $ the Wiener measure. $\pro B$ denotes the $3$-dimensional
Brownian motion on
$(\YYY , \mB(\YYY), \WW)$ starting at $0$. We denote by $\pro \FFF$ the natural
 filtration of the Brownian motion: $\FFF_t=\sigma(B_s, 0\leq s\leq t)$.

\bp{funda}
Assume (E1), (E2), (E3).
Then \kak{n1} has the unique solution $X^\rx=\pro{X^\rx}$ which is a diffusion process with respect to the filtration $\pro {{\FFF }}$:
\eq{ina}
\Ew{f(X_{s+t}^\rx)|{\mathcal F}_s}=
\Ew{f(X_t^{X_s^\rx})}
\en
for any bounded Borel measurable function $f$, where $\Ew{f(X_t^{X_s^\rx})}$ is
$\Ew{f(X_t^\ry)}$ evaluated at $\ry=X_s^\rx$.
\ep
\proof Since $b$, $\sigma$ are bounded and uniformly Lipschitz,
the proposition follows from \cite[Theorem 5.2.1]{Ok}. \qed

The following proposition is well-known. For lack of a precise reference, we will sketch its proof.
\bp{I10}
Assume (E1), (E2), (E3).
Then
\eq{yoko10}
\e^{-t\hz }f(\rx)= \Ew{f(X_{t}^{\rx})},\quad t\geq0,\hbox{ a.e. } \rx\in \rr^{3}.
\en
for $f\in\LR$.
\ep
\proof We first prove (\ref{yoko10}) for $f\in \coinf(\rr^{3})$, under the additional assumption that
\begin{equation}
\label{brinon}
\p_{\rx}^{\alpha}A^{jk}(\rx)\in L^{\infty}(\rr^{3}), \ \forall \ \alpha\in \nn^{3}.
\end{equation}
Let $u(t, \rx):= \e^{(t-T)K_0}f(\rx)$, $0\leq t\leq T$. By elliptic regularity we know that $\Dom K_{0}^{n}= H^{2n}(\rr^{3})$ and using that $u\in C^{k}([0, T], \Dom K_{0}^{n})$ we see using Sobolev's inequalities that $u(t, \rx)$ is a  bounded $C^{1, 2}([0, T]\times \rr^{3})$ solution of:
\[
\p_{t}u(t, \rx)= K_{0}u(t, \rx), \  u(T, \rx)= f(\rx).
\]
By \cite[Thm. 7.6]{ks} it follows that
\[
u(0, \rx)= \e^{- TK_{0}}f(\rx)= \Ew{f(X_{T}^{\rx})},
\]
which proves (\ref{yoko10}) in this case.

  We assume now only (E1), (E2), (E3).  We can find a sequence $[A^{jk}]_{n}(\rx)$ satisfying (\ref{brinon}),  such that $[A^{jk}]_{n}(\rx)$, $b_{n}(\rx)$ are uniformly Lipschitz and
  \[
  [A^{jk}]_{n}\to [A^{jk}], \ b_{n}\to b, \hbox{ uniformly in  }\rr^{3}.
\]
This also implies that $\sigma^{jk}_{n}\to \sigma^{jk}$ uniformly in $\rr^{3}$.

Let us denote by $X^{\rx}_{t, n}$ the solution of (\ref{n1}) with $\sigma_{n}$, $b_{n}$ and by $K_{0,n}$ the associated differential operator.
By a well known stability result for solutions of stochastic differential equations,  see e.g. \cite[Chapter 5]{E} we obtain that
\[
\Ew{|X^{\rx}_{t, n}- X^{\rx}_{t}|^{2}}\to 0, \ \forall \rx\in \rr^{3}.
\]
Let us fix $f\in \coinf(\rr^{3})$. Taking a sub sequence,  we obtain that $f(X^{\rx}_{t, n})\to f(X^{\rx}_{t})$ a.e. $\WW$ and hence that $T_{t, n}f(\rx)\to T_{t}f(\rx)$. On the other hand we see that $K_{0, n}\to K_{0}$ in norm resolvent sense, hence $\e^{- tK_{0, n}}f\to \e^{- tK_{0}}f$ in $L^{2}(\rr^{3})$. Taking again a	 sub sequence, we obtain that $\e^{- tK_{0,n}}f(\rx)\to \e^{- t K_{0}}f(\rx)$ a.e. $\rx$. Therefore the identity (\ref{yoko10}) holds for $f\in \coinf(\rr^{3})$, under assumptions (E1), (E2), (E3).

We first  extend (\ref{yoko10})   to $f\in L^{2}(\rr^{3})\cap L^{\infty}(\rr^{3})$ by density.
For $t\geq 0$, $f\in C_0^2(\BR)$ we set $$m_t(f)=
\int\EE_{\WW }[f(X_t^{\rx})]d\rx.$$
Clearly $f\geq0$ implies $m_t(f)\geq 0$. Since $K_0$ is a uniformly elliptic operator, $e^{-tK_0}$ is a contraction
on $L^1(\BR)$ \cite[Theorem 1.3.9]{Dav}.
Using again \kak{yoko10}
we get
\eq{ka1}
|m_t(f)|\leq \int_{\BR}|f(\rx)|d\rx, \ f\in C_0^2(\BR),
\en
and  \kak{ka1}
can be extended to $f\in L^1$.
It also follows from the Riesz-Markov theorem that there exists
a Borel measure
$\varrho_t $ on $\BR$
such that
\eq{g1}
\int_{\BR}f(\rx) d\varrho_t (\rx)=
m_t(f)
\en
for all $f\in C_0^2(\BR)$.
Together with \kak{ka1}
it follows that
\eq{ka2}
\left|
\int_{\BR}f(\rx) d\varrho_t (\rx)
\right|\leq
\int_{\BR} |f(\rx)|d\rx.
\en
Let now $f\in L^{2}(\rr^{3})\cap L^{\infty}(\rr^{3})$. We can find a sequence $(f_n)_{n\in\mathbb N}$ with $f_n\in C_0^2(\BR)$ such that
$f_n\to f$ in $L^2$, $f_n\to f$ a.e. in $\BR$ and $\sup_n\|f_n\|_\infty<\infty$.
Let us fix $t>0$. Since $f_n\to f$ in $L^2$, we get that $e^{-tK_0}f_n\to\e^{-tK_0}f$ in $\Dom  K_0=H^2(\BR)$,
 hence uniformly on $\BR$.
Let \[
\begin{array}{rl}
\ms N=&\{x\in\BR|f_n(x)\not\to
f(\rx)\}, \\[2mm]
\tilde{\ms N}=&
\{(\rx,\omega)\in \BR\times\mX_{+}|
f_n(X_t^{\rx}(\omega))\not \to
f(X_t^{\rx}(\omega))\}.
\end{array}
\]
By \kak{ka2} we have
\[
\begin{array}{rl}
&\int \one_{\tilde{\ms N}}
d\rx\otimes
d{\WW }
=
\int \one_{{\ms N}}(X_t^{\rx}(\omega))
 d\rx\otimes
d{\WW }\\[2mm]
&=
\int \one_{{\ms N}}(\rx)
d\varrho_t (\rx)
\leq
\int \one_{{\ms N}}(\rx)d\rx=0,
\end{array}
\]
since $f_{n}(\rx)\to f(\rx)$ a.e.
Hence
\eq{ka3}
f_n(X_t^{\rx}(\omega))\to f(X_t^{\rx}(\omega)),\hbox{ a.e. } (\rx,\omega)
\en
with respect to $d\rx\otimes
d{\WW }$.
Therefore using that $(f_n)_{n\in\mathbb N}$
is uniformly bounded, we have
$$\EE_{\WW }[f_n(X_t^{\rx})]
\to
\EE_{\WW }[f(X_t^{\rx})]\hbox{ a.e. } \rx,$$
which proves
(\ref{yoko10})  for
$f\in \LR\cap L^\infty(\BR)$.

Finally let us extend (\ref{yoko10})  to $f\in L^{2}(\rr^{3})$. We
may assume that $f\geq0$ without loss of generality.
We set $f_{n}:=\min\{f, n\}$, $n\in \nn$, so that $f_n\in\LR\cap L^\infty(\BR)$, $f_{n}(\rx)\nearrow f(\rx)$.
Since $\e^{-tK_0}$ is positivity preserving,
we see that
\[
(\e^{-tK_0}f_n)(\rx)
\nearrow (\e^{-tK_0}f)(\rx)<\infty, \hbox{ a.e. } \rx.
\]
By the same argument as above
we get
\[
f_{n}(X^{\rx}_{t}(\omega))\nearrow f(X^{\rx}_{t}(\omega)), \hbox{ a.e. }(\rx, \omega),
\]
and applying (\ref{yoko10})  to $f_{n}$  we see that $\sup_{n}\EE_{\WW }[f_n(X_t^{\rx})]<\infty$ a.e. $\rx$.
The monotone convergence theorem yields that
$\EE_{\WW }[f_n(X_t^{\rx})]\nearrow \EE_{\mathcal W}[f(X_t^{\rx})]<\infty$
 a.e. $\rx$, which completes the proof of the proposition.
\qed

\subsection{Feynman-Kac formula}\label{sec1.3b}

\bp{6}
{\bf (Feynman-Kac type formula)}
Let $f\in\LR$. Assume (E1), (E2), (E3), (E4).
Then
\eq{7}
\lk \e^{-tK }f\rk (\rx)={\EE}_{\WW}\lkkk
f(X_t^\rx)\e^{-\int_0^t \VV(X_s^\rx) ds}\rkkk.
\en
\ep
\proof
We assume for simplicity that $V$ is continuous. The extension to $V\in L^{1}_{\rm loc}$, $V\geq 0$ can be done by the same argument as in e.g. \cite[Thm. 6.2]{sim04}.
By the Trotter-Kato product formula \cite{km} we have
\eq{I14}
 \e^{-tK }f=
\lim_{n\to \infty}
(\e^{-(t/n)\hz}\e^{-(t/n)\VV})^n f.
\en
Let $0\leq s_{i}\in \rr$, $f_{i}\in L^{\infty}(\rr^{3})$ for $1\leq i \leq n$. By Proposition  \ref{I10} we have:
\begin{align}
&
 \lk
 \e^{-s_{1}\hz}f_1\cdots \e^{-s_{n}\hz}f_n
 \rk(\rx)\non \\
&=
\Ew{f_1(X_{s_{1}}^\rx)
\lk
\e^{-s_{2}\hz}f_1\cdots \e^{-s_{n}\hz}f_n\rk(X_{s_{1}}^\rx)}\non\\
&=
\Ew
{
f_1(X_{s_{1}}^\rx)
\Ew
{
f_2(X_{s_{2}}^{X_{s_{1}}^\rx})
\lk
\e^{-s_{3}\hz}f_3\cdots \e^{-s_{n}\hz}f_n\rk(X_{s_{2}}^{X_{s_{1}}^\rx})
}}.
\non\end{align}
By the Markov property \kak{ina} we also have
\begin{align*}
 &
\Ew
{
f_1(X_{s_{1}}^\rx)
\Ew
{
f_2(X_{s_{2}}^{X_{s_{1}}^\rx})
\lk
\e^{-s_{3}\hz}f_3\cdots \e^{-s_{n}\hz}f_n\rk(X_{s_{2}}^{X_{s_{1}}^\rx})
}}\\
&=
\Ew{
f_1(X_{s_{1}}^\rx)
\Ew{
f_2(X_{s_{1}+ s_{2}}^\rx)
\lk
\e^{-s_{3}\hz}f_3\cdots \e^{-s_{n}\hz}f_n\rk(X_{s_{1}+ s_{2}}^\rx)\left|
\FFF _{s_{1}}\right.
}}\\
&=\Ew{
f_1(X_{s_{1}}^\rx)
f_2(X_{s_{1}+ s_{2}}^\rx)
\lk
\e^{-s_{3}\hz}f_3\cdots \e^{-s_{n}\hz}f_n\rk(X_{s_{1}+ s_{2}}^\rx)
}.
\end{align*}
Inductively we obtain that
\eq{I13}
 \lk
 \e^{-s_{1}\hz}f_1\cdots \e^{-s_{n}\hz}f_n
 \rk(\rx)
=\Ew{
\prod_{j=1}^n f_j(X_{t_j}^\rx)}, \hbox{ for } t_{1}= s_{1}, \ t_{j}= t_{j-1}+ s_{j}.
\en

Combining the Trotter product formula \kak{I14} and \kak{I13} with $s_j=t/n$, $f_{j}= \e^{- (t/n)V}$
we have
\eq{I15}
 \e^{-tK }f(\rx)=
\lim_{n\to \infty}
\Ew{
\e^{-(t/n)\sum_{j=1}^ n\VV(X_{tj/n}^\rx)}
f(X_t^\rx)
}.
\en
Since $t\to X_t^{\rx}$ is continuous a.e. $\WW$ and $\VV$ is continuous it follows that
\[
(t/n)\sum_{j=1}^ n\VV(X_{tj/n}^\rx)\to \int_{0}^{t}\VV(X^{\rx}_{s}) ds,\hbox{ a.e. }\WW\hbox{ when }n\to \infty.
\]
Using that $\VV(\rx)\geq 0$ and the Lebesgue dominated convergence theorem, we obtain that
\[
\Ew{
\e^{-(t/n)\sum_{j=1}^ n\VV(X_{tj/n}^\rx)}
f(X_t^\rx)
}\to \Ew{
\e^{-\int_{0}^{t}\VV(X_{s}^{\rx}ds)}
f(X_t^\rx)
}\hbox{ in }L^{2}(\rr^{3}).
\]
This completes the proof of the proposition. \qed

\subsection{Bounds on heat kernels}
We first recall some easy consequences of the Feynman-Kac formula.
\bp{zang}
Assume (E1), (E2), (E3), (E4).
Then there exist constants $C, c>0$ such that
\eq{yu2}
 \e^{-tK}(\rx,\ry)
\leq
c \e^{Ct\Delta}(\rx,\ry), \ t\geq 0, \ \hbox{ a.e. }\rx, \ry\in \rr^{3}.
\en
Here
$$
\e^{T \Delta}(\rx,\ry)=
(4\pi T)^{-3/2}
\e^{-|\rx-\ry|^2/(4T)}$$ is the three dimensional heat kernel.
\ep
\proof By the Feynman-Kac formula we know that
\[
\e^{-tK}(\rx,\ry)\leq \e^{- tK_{0}}(\rx, \ry),\quad t\geq 0, \ \hbox{ a.e. }\rx, \ry\in \rr^{3}.
\]
Then we apply Proposition \ref{irot}.
 \qed

\medskip

Using the upper bound in Proposition \ref{zang}, we get the following corollary.
\bc{ultra}{\bf (Ultracontractivity)}
 Assume (E1), (E2), (E3), (E4). Then $\e^{-tK }$ maps $\LR$ to $L^\infty(\BR)$ for $t>0$.
\ec

\bc{pi}{\bf (Positivity improving)}
Assume (E1), (E2), (E3), (E4).
Then $\e^{-tK}$ is positivity improving for $t>0$.
In particular if (E5) holds $K$ has a unique strictly positive ground state.
\ec
\proof

We first claim that
\begin{equation}
\label{intV}
\int_{0}^{t}V(X_{s}^{\rx}) ds<\infty, \hbox{ a.e.} (x, \omega).
\end{equation}
Assume first that $V\in L^{1}(\rr^{3})$. Then since $\e^{- sK}$  are contractions on $L^{1}$ we get:
\[
\int_{\rr^{3}} d\rx  \Ew{\int_{0}^{t}V(X^{\rx}_{s}) ds}= \int_{0}^{t}(1| \e^{- s K}V) ds\leq t \|V\|_{1},
\]
hence (\ref{intV}) holds for $V\in L^{1}(\rr^{3})$. If $V\in L^{1}_{\rm loc}(\rr^{3}$, then $V_{n}:= \one_{\{|x|\leq n\}}V\in L^{1}(\rr^{3})$ and there exist
sets ${\mathcal N}_{n}\in \rr^{3}\times \YYY$ of measure zero such that
\[
\int_{0}^{t}V_{n}(X_{s}^{\rx}) ds<\infty, \ (x, \omega)\in \mathcal{ N}_{n}.
\]
Set $\mathcal{ N}:= \bigcup_{n\geq 1}\mathcal{ N}_{n}$.  Since $s\mapsto X_{s}^{\rx}(\omega)$ is continuous, for each $(\rx, \omega)$ there exists $N=N(\rx, \omega)\in \nn$ such that $N\geq \sup_{0\leq s \leq t}|X^{\rx}_{s}(\omega)|$ and hence $V(X_{s}^{\rx}(\omega))= V_{N}(X_{s}^{\rx}(\omega))$ for all $0\leq s \leq t$. Therefore
\[
\int_{0}^{t}V(X_{s}^{\rx}) ds<\infty, (x, \omega)\not\in \mathcal{ N},
\]
which proves (\ref{intV}). To prove that $\e^{- tK}$ is positivity improving it suffices to prove that for $f, g\geq 0$ with $f, g\not\equiv 0$ one has $(f|e^{- tK}g)>0$. Assume that
\beq\label{orlov}
(f|\e^{- tK}g)= \int_{\rr^{3}}d\rx\Ew{f(\rx)g(X^{\rx}_{t})\e^{-\int_{0}^{t}V(X^{\rx}_{s})ds}}=0.
\eeq
It follows from (\ref{intV}) that $\e^{-\int_{0}^{t}V(X^{\rx}_{s})ds}>0$ a.e. $(\rx, \omega)$. Hence (\ref{orlov}) implies that
\[
\int_{\rr^{3}}d\rx\Ew{f(\rx)g(X^{\rx}_{t})}=(f|e^{- tK_{0}}g)=0.
\]
But this contradicts the lower bound in Prop. \ref{irot}. \qed

\medskip

\begin{lemma}\label{ex2}{\bf (Exponential decay)}
Assume (E1), (E2), (E3), (E5). Let $\grp$ be the unique strictly positive ground state of $K$.
Then there $\delta>0$ such that
\[
\e^{ \delta | \rx|^{\delta+1}}\grp\in H^{1}(\rr^{3}).
\]
\end{lemma}
\proof
If $F\in C^{\infty}(\rr^{3})$ is real, bounded with all derivatives, then for $u\in \Dom K$ we have the well-known Agmon identity:
\begin{eqnarray*}
&&\int \12\langle \nabla(\e^{F}\overline{u}),A\nabla(\e^{F}u)\rangle d\rx + \int\e^{2F}(\VV-\12 \langle \nabla F, A\nabla F\rangle)|u|^{2} d\rx\\
&&=
\int \e^{2F}\overline{u}Ku d\rx +2\i {\rm Im}\int \e^{2F}\langle \nabla\overline{u},A\nabla F\rangle ud\rx.
\end{eqnarray*}
Applying this identity to the real function $\grp$, we obtain by the usual argument that there exists $\delta>0$ such that $\e^{ \delta\langle \rx\rangle^{\delta+1}}\grp\in L^{2}(\rr^{3})$ and $\nabla(\e^{ \delta\langle \rx\rangle^{\delta+1}}\grp)\in L^{2}(\rr^{3})$. \qed

\subsection{Ground state transformation and diffusion process}

Assume (E1), (E2), (E3) and (E5). Then $K$ has compact resolvent and by Corollary \ref{pi} it has a unique normalized strictly positive ground state $\grp $. We set
\eq{vo1}
\mup=\grp^2(\rx) d\rx, \ \HP=L^2(\BR,d \mu_{\rm p}),
\en
and introduce the {\em ground state transformation:}
$$\mU_p:\HP\to \LR,\quad f\mapsto \grp f.$$
Let $\lp$ be
the corresponding transform of $K -\is({K })$ defined by
\eq{y1-2}
\lp=\mU_p \f (K -\is({K })) \mU_p
\en
with $\Dom (\lp)= \mU_p\f \Dom (K)$. We note that
$$(f| \lp g)_{\HP}=(\grp f| K \grp g)_{L^2}-
\is(K)(\grp f,|\grp g)_{L^2}
.$$
Our goal in this subsection is to construct a
 three dimensional {\em diffusion process} (i.e., a continuous Markov process)
 $X=(X_t)_{t\in\rr} $
associated with  $\lp$.  The operator $\lp$ is
formally
of the form
\eq{sa1}
\lp=
-\half \sum_{1\leq j,k\leq 3}
A^{jk}
\p_{\rx_j}
\p_{\rx_k}
+
\sum_{1\leq j,k\leq3} \lk
\half (\p_{\rx_j} A^{jk})+
A^{jk}
\frac{\p_{\rx_j} \grp}{\grp}
\rk
\p_{\rx_k}.
\en
A standard way
to construct
the diffusion process $X_{t}$
is to solve
the following stochastic differential
equation:
\eq{sde}
dX_t^j= \sum_{k=1}^3\sigma^{jk}(X_t) dB_t^k+
\sum_{k=1}^3
\lk
\half (\p_k A^{jk})
(X_t)
+
A^{jk}(X_t)
\frac{\p_k \grp(X_t)}{\grp(X_t)}
\rk dt
\en
derived from
\kak{sa1},
where $B_t$ denotes
the three-dimensional Brownian motion,
and the diffusion term is $\sigma(\rx)=[A^{jk}]^{\12}(\rx)$.
This is
of course a formal description, since the regularity of $\grp$
is not clear at all, and it is thus
hopeless
to solve \kak{sde} directly.
Instead of this direct approach
we use another strategy to construct
the diffusion process $X$ associated with
$\lp$.
This  is done in Appendix \ref{a4}.

 We summarize the properties of $X_{t}$ in Proposition  \ref{main1} below. Let
$\mX =C(\rr ;\BR)$.
 $X\stackrel{\rm d}{=}
Y$ means that $X$ and $Y$
has the same distribution.
\begin{proposition}\label{main1}
{\bf (Diffusion process associated with $\e^{-t\lp}$)}
Let
$$X_t(\omega)=\omega(t),\quad \omega(\cdot)\in \mX,$$
be the coordinate mapping process
on $(\mX , \mB(\mX))$, where $\mB(\mX)$ denotes
the $\sigma-$field generated
by cylinder sets.
Assume (E1), (E2), (E3), (E5).
Then there exists for all $\rx\in \rr^{3}$
a probability measure ${\rP^{\rx}}$ on $(\mX ,
\mB(\mX))$
satisfying (1)-(5) below:
\bi
\item[(1)]
{\bf (Initial distribution)}
 $\rP^\rx(X_0=\rx)=1$.
 \item[(2)]
 {\bf (Continuity)} $t\mapsto X_{t}$ is continuous.
\item[(3)]
{\bf (Reflection symmetry)}
$\pro{X}$ and $(X_t)_{t\leq0}$
are independent\\
 and $X_{-t}\stackrel{\rm d}{=}
 X_t$.
\item
[(4)]
{\bf (Markov property)}
Let
$(\FFF _t)_{t\geq 0}=\sigma(X_s, 0\leq s\leq t)$ for $t\geq 0$
and
$(\FFF _t)_{t\leq 0}=\sigma(X_s, t\leq s\leq 0)$ for $t\leq 0$ be the associated filtrations.
Then
$\pro{X}$ and $(X_t)_{t\leq 0}$
are Markov processes with respect to $\pro {\FFF }$ and $(\FFF _t)_{t\leq 0}$, respectively,
 i.e.
\eqn
&&
\EE_{\rP^\rx}[X_{t+s}|\FFF _s]=
\EE_{{\rP^{\rx}}}[X_{t+s}|\sigma(X_s)]
=
\EE_{\rP^{X_s}}[X^{X_s}_{t}],\\
&&
\EE_{\rP^\rx}[X_{-t-s}|
\FFF _{-s}]=
\EE_{\rP^\rx}[X_{-t-s}|\sigma(X_{-s})]
=\EE_{\rP^{X_{-s}}}[X^{X_{-s}}_{-t}]
\enn
for $s,t\geq0$, where
$\EE_{\rP^{X_s}}$ means $\EE_{\rP^{\ry}}$
evaluated at $\ry=X_s$.
\item[(5)]{\bf (Shift invariance)}
\eq{I4-1ko}
\int\mup
\EE_{\rP^\rx}
\lkkk
 {f_0(X_{t_0})\cdots f_n(X_{t_n})}\rkkk
=(f_0|\e^{-(t_1-t_0)\lp}f_1\cdots \e^{-(t_n-t_{n-1})\lp} f_n)_{\HP}
\en
for $f_j\in L^\infty(\BR)$, $j=1,...,n$, and the finite dimensional distribution of $X$
is shift invariant, i.e.:
$$
\int\mup
\EE_{\rP^\rx}
\lkkk
\prod_{j=1}^n f_j(X_{t_j})\rkkk
=
\int\mup
\EE_{\rP^\rx}
\lkkk
\prod_{j=1}^n f_j(X_{t_j+s})\rkkk,
\quad
s\in\rr ,
$$ for all bounded Borel measurable functions $f_j$, $j=1,...,n$.
\ei
\end{proposition}
This proposition may be known,
the proof will be however given in Appendix \ref{a4} for self consistency.
We define the full probability measure
$\rP$ on $\BR\times \mX$
by
$$\rP(A\times B)=
\int _{A} d\mu_{\rm p}(\rx)
\int_{B} d{\rP^{\rx}}.$$
In the sequel we will denote $\EE_{\rP^\rx}$ simply by $\EE^\rx$.
\section{The Nelson model by path measures}\init\label{sec2}

\subsection{Path space approach for boson fields}
Let $\cX$ be a real Hilbert space and $a> 0$ a selfadjoint operator
on $\cX$. It is well known that there exist a
probability space $(Q, \Sigma, \mu_{C})$ and a linear map:
\[
a^{\12}\cX\ni f\mapsto \Phi(f)
\]
with values in measurable functions on $(Q, \Sigma)$ such that
\[
\int_{Q}\e^{\i \Phi(f)}d\mu_{C}= \e^{-\12C(f,f)}, \ f\in
a^{\12}\cX,
\]
for $C(f,f)= (f|a^{-1}f)_{\cX}$.
Moreover $\Sigma$ is generated by the functions $\Phi(f)$, $f\in
a^{\12}\cX$.

Such a structure is called the {\em Gaussian process} indexed by $\cX$ with
{\em covariance} $C$.
Let $\cX_{\cc}$ be the complexification of $\cX$.
It is well known that $L^{2}(Q, d\mu_{C})$ can be unitarily
identified with the bosonic Fock space $\Gamma_{\rm s}(a^{\12}\cX_{\cc})$ by the
map
\beq\label{mappo}
U: L^{2}(Q, d\mu_{C})\ni \e^{\i \Phi(f)}
 \mapsto \e^{\i\phi(f)}\Omega\in \Gamma_{\s}(a^{\12}\cX_{\cc}),
 \ \ f\in a^{\12}\cX.
\eeq
Here we recall that $\Omega$ is the Fock vacuum.
If we further identify
$\Gamma_{\rm s}(a^{\12}\cX_{\cc})$ with
$\Gamma_{\rm s}(\cX_{\cc})$ by the map $\Gamma(a^{-\12})$, we obtain
that $\Gamma_{\rm s}(\cX_{\cc})$ is unitarily identified with $L^{2}(Q,
 d\mu_{C})$ by
 $\mU_{\rm f}= \Gamma(a^{-\12})U$:
\eq{42}
\mU_{\rm f}:
L^2(Q,d\mu_C)\ni \e^{\i \Phi(f)}
\mapsto
\e^{\i \phi(a^{-\12}f)}\Omega\in
\Gamma_\s(\cX_{\cc})
, \ f\in a^{\12}\cX.
\en

We will apply this result to $\cX= \LR$ and
$a=2\omega$, where $\omega$ is defined in (\ref{ash}) (note that
$\omega$ is a real operator). The associated probability space will
be denoted by $(Q_{0}, \Sigma_{0}, \mu_{0})$ and we set
$$
\cH_{\rm f}:=L^{2}(Q_{0}, d\mu_{0}).
$$
Note that under the above identification, any closed operator $T$ on
$\Gamma_{\rm s}(L^{2}(\rr^{3}))$ affiliated to the abelian von Neumann
algebra generated by the $\e^{\i \phi(g)}$ for $g\in
L^{2}_{\rr}(\rr^{3})$ becomes a multiplication operator by a
measurable function  on
$(Q_{0}, \Sigma_{0})$. We set
\[
H_{\rm f}:= \mU_{\rm f}^{-1} d\Gamma(\omega)\mU_{\rm f}.
\]

We now recall the well known expression of the semi-group $\e^{-
tH_{\rm f}}$ through Gaussian processes.
Let us set $D_t=-\i\p_t$.
Consider the Gaussian process indexed by $L_{\rr}^{2}(\rr^{4})=
L_{\rr}^{2}(\rr, dt)\otimes L^{2}_{\rr}(\rr^{3}, dx)$
 with covariance
 \eq{cov}
 C(f, f)= (f| (D_{t}^{2}+\omega^{2})^{-1}f)
 _{L^2(\rr^4)}
 \en
 and set $a= D_{t}^{2}+ \omega^{2}$. The
associated probability space will be denoted by \\ $(Q_{\rm E}, \Sigma_{\rm E}, \mu_{\rm E})$ and the random variables by
$\Phi_{\rm E}(f)$.
It is well known that for $t\in \rr$, the map
\[
j_{t}: (2\omega)^{\12}L^{2}(\rr^{3})\ni g\mapsto
\delta_{t}\otimes g\in a^{\12}L^{2}(\rr^{4})
\]
is isometric, if $\delta_{t}$ denotes the Dirac mass at time $t$.
Moreover one has
\beq\label{ne.e1}
(\delta_{t_{1}}\otimes g_{1}|(D_{t}^{2}+ \omega^{2})^{-1}
\delta_{t_{2}}\otimes g_{2})_{L^{2}(\rr^{4})}=
(g_{1}|\frac{1}{2\omega}\e^{-|t_{1}- t_{2}|
\omega}g_{2})_{L^{2}(\rr^{3})}.
\eeq
For $t\in \rr$ and $g\in
(2\omega)^{\12}L^{2}_{\rr}(\rr^{3})$, we set
$$\Phi_{\rm E}(t, g):=
\Phi_{\rm E}(\delta_{t}\otimes g).$$
It
follows that
  $\Phi_{\rm E}(t, g)\in \cap_{1\leq p<\infty}L^{p}(Q_{\rm E}, {\rm d}\mu_{\rm E})$.
Since the covariance $C$ is invariant under the group $\tau_{s}$ of time
translations, we see that
$$T_{s}:= \mU^{-1}_{\rm f}\Gamma(\tau_{s})\mU_{\rm f},\quad s\in\rr$$ is a strongly continuous unitary
group on $L^{2}(Q_{\rm E}, d\mu_{\rm E})$.
Clearly
\[
T_{s}\Phi_{\rm E}(t, g)= \Phi_{\rm E}(t+s, g).
\]
For $t\in \rr$, we denote by $E_{t}: L^{2}(Q_{\rm E}, \Sigma_{\rm E}, \mu_{\rm E})\to L^{2}(Q_{\rm E}, \Sigma_{\rm E}, \mu_{\rm E})$
 the conditional expectation
with respect to the $\sigma$-algebra $\Sigma_{t}$ generated by the
$\Phi(t, g)$ for $g\in (2\omega)^{\12}L^{2}_{\rr}(\rr^{3})$. As is well known one has
$E_{t}=
\mU_{\rm f}^{-1} \Gamma(e_t)\mU_{\rm f}$
for
$e_{t}= j_{t}j_{t}^{*}$.
Clearly $E_{0}L^{2}(Q_{\rm E}, d\mu_{\rm E})$ can be identified with
 $\cH_{\rm f}$
and we will hence consider
$\cH_{\rm f}$ as a closed subspace of $L^{2}(Q_{\rm E}, d\mu_{\rm E})$.
 It follows then from
(\ref{ne.e1}) that
\beq\label{19}
\int_{Q_{\rm E}} \overline{F}T_{t}G d\mu_{\rm E}= (F|\e^{-
t \hf}G)_{\cH_{\rm f}}, \ t\geq 0,
\eeq
for $F, G\in \cH_{\rm f}$.

\subsection{Path space representation for the Nelson model}
The Hilbert space $\cH_{\rm p}\otimes \cH_{\rm f}\cong L^2(\rr^3\times Q_0,d\mu_{\rm p}\otimes d\mu_0)$ and the Hamiltonian $(\mU_{\rm p}\otimes
\mU_{\rm f})H(\mU_{\rm p}\otimes \mU_{\rm f})^{-1}$ will still be denoted by $\cH$ and
$H$ respectively.
Note that $F\in \cH$ can be viewed
as a function: $F: \rr^{3}\ni \rx\mapsto F(\rx)\in \cH_{\rm f}$ defined
almost everywhere. Note also that in this representation the interaction $q\varphi_{\rho}(\rx)$ becomes the multiplication operator by the measurable function on $Q_{\rm E}\times \rr^{3}$:
$q\Phi_{\rm E}(0, \rho(\cdot -\rx))$.
\begin{theoreme}\label{irto}
\ben

\item Assume (E1), (E2), (E3), (E5). Let $F,G\in \cH$.
 Then for all $t\geq 0$
\beq\label{fkn-formula}
(F|\e^{-tH}G)_{\cH}=
\int d\mu_{\rm p}(\rx){\EE}^{\rx}\left[(F(X_{0})|\e^{-q\int_{0}^{t} \Phi_{\rm E}(s, \rho(\cdot -X_{s})) d
s}T_{t}G(X_{t}))_{L^{2}(Q_{\rm E})} \right].
\eeq
\item In particular \eq{I20}
(\one| \e^{-TH}\one)_\hhh=
\ix {
\e^{(q^{2}/2) \int_0^T dt \int_0^T ds
W( X_t, X_s, |t-s|)}},
\en
where
\eq{num}
W(\rx,\ry, |t|)=
(\rho(\cdot-\rx),
\frac{\e^{-|t|\omega}}{2\omega}
\rho(\cdot-\ry))_\LR.
\en
\een
\end{theoreme}

\proof
Suppose that
$G\in L^\infty(\BR\times Q_{0},d\mu_{\rm p}\otimes d\mu_0)$.
By the Trotter-Kato product formula \cite{km} we have
$$\e^{-tH}=
s-\lim_{n\rightarrow\infty}
\lk \e^{-(t/n)\lp }\e^{-(t/n)
 \Phi_{\rm E}(0, q\rho(\cdot -\rx))}
\e^{-(t/n)\hf}\rk^n.
$$
Using the factorization formula \kak{19}, the Markov property of $E_t$,
 we have
\begin{align}\label{yoko3}
&
(F|\e^{-tH}G)\\
&=
\lim_{n\rightarrow \infty}
\int \mup \EE^\rx\left[
\lk
F(X_0),
\e^{-
\sum_{j=0}^n
\frac{t}{n}
\Phi_{\rm E}(jt/n, q\rho(\cdot -X_{jt/n}))}
T_{t}G(X_t)
\rk
\right].\notag
\end{align}
We set
$\rho_{s}= \rho(\cdot - X_{s})$ for $s\in \rr$.
Using that $s\mapsto X_{s}\in \rr^{3}$ is continuous, we see that
\beq\label{koko}
\rr\ni s\mapsto \rho_{s}\in \omega^{\12}L^{2}(\rr^{3})
\eeq
is continuous.
This implies that
 \beq\label{koko2}
\rr\ni s \mapsto \Phi_{\rm E}(s, \rho_{s})\in \cap_{1\leq p<\infty}L^{p}(Q_{\rm E}, d \mu_{\rm E})
\eeq
 is also continuous.
In fact
\[
\Phi_{\rm E}(t, \rho_{t})- \Phi_{\rm E}(s, \rho_{s})= \Phi_{\rm E}(t, \rho_{t})- \Phi_{\rm E}(s, \rho_{t})+ \Phi_{\rm E}(s, \rho_{t}- \rho_{s}).
\]
The first term in the right hand side tends to $0$ when $s\to t$ in $\cap_{1\leq p<\infty}L^{p}(Q_{\rm E}, d \mu_{\rm E})$ since the time translation group
$\{T_{t}\}_{t\in \rr}$ is strongly continuous \\ on $\cap_{1\leq p<\infty}L^{p}(Q_{\rm E}, d \mu_{\rm E})$. The same is true for the second term using (\ref{koko}).
It follows from (\ref{koko2}) that
\[
\sum_{j=0}^n
\frac{t}{n}
\Phi_{\rm E}(jt/n, \rho(\cdot -X_{jt/n}))\to \int_{0}^{t}\Phi_{\rm E}(s, \rho(\cdot - X_{s})) ds
\]
in
  $\cap_{1\leq p<\infty}L^{p}
  (Q_{\rm E})$,
when $n\to \infty$.
We claim now that
\begin{equation}
\label{kiki}
\lim_{n\to +\infty}
\exp\lk
-
\sum_{j=0}^n
\frac{t}{n}
\Phi_{\rm E}(jt/n, q\rho(\cdot -X_{jt/n}))
\rk
= \exp\lk
-q\int_{0}^{t} \Phi_{\rm E}(s, \rho(\cdot -X_{s}))d
s\rk
\end{equation}
in
  $\cap_{1\leq p<\infty}L^{p}
  (Q_{\rm E})$.
To prove (\ref{kiki}) we set
\[
\begin{array}{l}
\d f_{n}= (t/n)\sum_{j=1}^{n}\delta_{tj/n}\otimes \rho(\cdot - X_{tj/n}(\omega)), \\[2mm]
\d f= \int_0^t
\delta_s
\otimes \rho(\cdot-X_s) ds.
\end{array}
\]
The map
$$\rr\ni s\mapsto
\delta_s
\otimes \rho(\cdot-X_s)\in (D_t^2+\omega^2)^{\12}L^2(\rr^4)$$
is continuous.
Note that
\[
\begin{array}{l}
\d \Phi_{\rm E}(f_{n})= \sum_{j=0}^n
\frac{t}{n}
\Phi_{\rm E}(jt/n, \rho(\cdot -X_{jt/n}))\\[2mm]
\d \Phi_{\rm E}(f)=
\int_{0}^{t} \Phi_{\rm E}(s, \rho(\cdot -X_{s}))ds.
\end{array}
\]
We write now
 \[
\e^{-q\Phi_{\rm E}(f)}-\e^{-q\Phi_{\rm E}(f_{n})}= q\int_{0}^{1}
\e^{-q\Phi_{\rm E}
(rf+ (1-r)f_{n})
}
\Phi_{\rm E}(f_{n}- f)dr.
\]
We use the fact that $\Phi_{\rm E}(g)$ is a Gaussian random variable and hence
\beq\label{koko3}
\|\e^{\Phi_{\rm E}(g)}\|_{L^{p}}^p
= \e^{p^2 C(g, g)/2}= \e^{(p^2/2)\|
\Phi_{\rm E}(g)\|_{2}^{2}}, \ 1\leq p<\infty, \ g \in
(D_{t}^{2}+ \omega^{2})^{\12}L^{2}(\rr^{4}).
\eeq
Applying H\"{o}lder's inequality we obtain (\ref{kiki}). To complete the proof of (1) it remains to justify the exchange of limit and integral. To do this we note that the family of functions
\[
F(X_{0})\e^{- q\Phi_{\rm E}(f_{n})}T_{t}G(X_{t}), \ n\in \nn
\]
is equi-integrable if $F\in L^{\infty}$, $G\in \cH$, since it is uniformly bounded in $L^{p}$ for some $p>1$, by H\"{o}lder's inequality and (\ref{kiki}). This completes the proof of (1) for $G\in L^\infty$ and $F\in \cH$.

Next suppose that $G,F\in\cH$.
We can suppose $F,G\geq0$ without loss of generality. Let $G_{n}=\min\{G, n\}$, $n\in \nn$.
 Thus $(F| \e^{-tH}G_n)\to(F| \e^{-tH}G)$ as $n\to\infty$
and
$
F(X_{0})\e^{-q\int_{0}^{t} \Phi_{\rm E}(s, \rho(\cdot -X_{s})) d
s}T_{t}G_n(X_{t})
$ is monotonously increasing as
$N\uparrow\infty$.
By the monotone convergence theorem we get  that $$F(X_{0})\e^{-q\int_{0}^{t} \Phi_{\rm E}(s, \rho(\cdot -X_{s})) d
s}T_{t}G(X_{t})\in
L^1(\BR\times {\mX }\times Q_{\rm E},
d\rP\otimes d\mu_{\rm E})$$ and (1) follows.
Applying (1) to $F= G=\one$, we get
$$(\one|\e^{-tH}\one)
=
\int\mup\EE^\rx\lkkk
(\one, \e^{q\Phi_{\rm E}(f)}\one)\rkkk
=
\int\mup\EE^\rx \lkkk
\e^{(q^{2}/2)C(f, f)}
\rkkk.
$$
Using  \kak{cov},
 we get
\begin{align*}
C(f, f)&=
\int_0^T dt \int_0^T ds
(\delta_t\otimes \rho(\cdot-X_t)|
(D_t^2+\omega^2)^{-1}
\delta_s\otimes \rho(\cdot-X_s)
)\\
&=
\int_0^T dt \int_0^T ds
W( X_t, X_s, |t-s|).
\end{align*}
This completes the proof of the theorem.
\qed

\medskip

\begin{proposition}\label{ircam}
Assume (E1), (E2), (E3), (E5). Then $\e^{-tH}$ is positivity improving for all $t>0$.
\end{proposition}
\proof Let $t>0$ and $F, G\in \cH$ with $F, G\geq 0$, $F, G\neq 0$.
We need to prove that $(F|\e^{-tH}G)>0$.
Since $\int_{0}^{t}\Phi_{\rm E}(s, \rho(\cdot - X_{s})) ds$ belongs to $L^{1}$, $\e^{- \int_{0}^{t}\Phi_{\rm E}(s, \rho(\cdot - X_{s})) ds }>0$
a.e. Therefore it suffices to prove that
\beq\label{argh}
\int d\mu_{\rm p}(\rx)
{\EE}^{\rx}\left[(F(X_{0})|T_{t}G(X_{t}))
\right]
= (F|\e^{- tH_{0}}G)>0.
\eeq
The equality above
immediately shows that $\e^{- tH_{0}}$ is positivity preserving for all $t>0$. Moreover $\one\otimes\one$ is the unique strictly positive ground state of $H_{0}$. Therefore by \cite[Theorem XIII.44]{rs4} $\e^{-tH_{0}}$ is positivity improving for all $t>0$ and hence (\ref{argh}) holds. This completes the proof of the proposition. \qed

\medskip

We complete this section by stating a standard abstract criterion for the existence of a ground state for generators of positivity improving heat semi-groups.

\begin{lemma}\label{auxili}
Let $(Q, \Sigma, \mu)$ be a probability space, and $H$ a bounded below selfadjoint
operator on $L^{2}(Q, \Sigma, \mu)$ such that $\e^{-tH}$ is positivity improving for
$t>0$. Set
\[
\gamma(T):= \frac{(\one|
\e^{- TH}\one)^{2}}
{\|\e^{-TH}\one\|^{2}},
\]
and $E= \inf \sigma(H)$.
Then $\lim_{T\to +\infty}\gamma(T)= \|\one_{\{E\}}(H)\one\|^{2}$.
In particular $H$ has a ground state iff
$\lim_{T\to+\infty}\gamma(T)\neq 0$.
\end{lemma}
 Note that by Proposition \ref{ircam},
 Lemma \ref{auxili} can be applied to the Nelson Hamiltonian $H$.

\proof
We can assume that $E=0$, so that $\slim_{T\to
+\infty}\e^{-TH}= \one_{\{0\}}(H)$. If $0$ is an eigenvalue, then by
Perron-Frobenius arguments,
$\one_{\{0\}}(H)= |u)(u|$ for some $u>0$. It follows that
$\lim_{T\to +\infty}\gamma(T)= (u|1)^{2}$.
Assume now that $H$ has no ground state and that there exists a
sequence $T_{n}\to +\infty$ such that $\gamma(T_{n})\geq \delta^{2}>0$.
This implies that
$(\one |\e^{-T_{n}H} \one)\geq
\delta(\one|\e^{-2T_{n}H}\one)^{\12}$. Letting $n\to +\infty$, we obtain
that $\|\one_{\{0\}}(H)\one\|\geq \delta$, which is a contradiction.
\qed

\section{Absence of ground state}
\subsection{Proof of Theorem \ref{hiroshima4}}
\init\label{sec4}
In this section we assume the hypotheses of Theorem \ref{hiroshima4}. We first prove some upper and lower bounds on the interaction kernels $W(\rx, \ry, t)$. This is the only place where the hypotheses
(B2) on fast decay of the variable mass $m(x)$ and (B3) on the positivity of the space cutoff function $\rho$ enter.

Set $\rho_\rx(x)=\rho(x-\rx)$.
We recall from (\ref{num}) that
\[
W(\rx, \ry, t)= (\rho_{\rx}|\frac{\e^{- t\omega}}{2\omega}\rho_{\ry}),
\ \rx, \ry\in \rr^{3}, \ t\geq 0.
\]
We set $h_{\infty}= -\Delta_{x}$, $\omega_{\infty}= h_{\infty}^{\12}$,
and denote by $W_{\infty}(\rx, \ry, t)$ the analog potential for
$\omega$ replaced by $\omega_{\infty}$.
Note also that
\[
\e^{-t\omega_{\infty}}(x, y)= \frac{1}{\pi^{2}}\frac{t}{(|x-y|^{2}+
|t|^{2})^{2}},
\]
which using the identity $\frac{1}{\lambda}\e^{- t\lambda}=\int_{t}^{+\infty}\e^{-
s \lambda}d s$ yields
\begin{eqnarray}
W_{\infty}(\rx, \ry, t)
&=&
 \frac{1}{4\pi^{2}}\int
\frac{\rho(x-\rx)\rho(y- \ry)}{|x-y|^{2}+ t^{2}}d xd y\nonumber\\
&=&
\frac{1}{4\pi^{2}}\int
\frac{\rho(x)\rho(y)}{|x-y+ \rx-\ry|^{2}+ t^{2}}d xd y.
\label{ne.e6}
\end{eqnarray}
We also have
\begin{equation}
\label{ne.e6b}
W_\infty(\rx,\ry,|t|)=\half \int\frac{|\hat\rho|^2(k)
\e^{-ik\cdot(\rx-\ry)}}{|k|}\e^{-|t||k|}\ dk.
\end{equation}

\bl{suzuki}
\ben
\item $W(\rx,\ry,|t|)\geq 0$ and $W_\infty(\rx,\ry,|t|)\geq 0$,
\item Assume (B2). Then there exist constants $C_j>0$, $j=1,2,3,4$,
 such that
\eq{ko2}
C_1 W_\infty(\rx,\ry,C_2|t|)
\leq W(\rx,\ry,|t|)\leq
C_3 W_\infty(\rx,\ry,C_4|t|)
\en
for all $\rx,\ry\in \BR$ and $t\in\rr $.
\een
\el
\proof
We note that the function
$f(\lambda)=\e^{-\sqrt \lambda}$ on $[0,\infty)$ is completely monotone, i.e.,
$(-1)^n df(\lambda)/d\lambda^n\geq0$ and that $f(+0)=0$. Then by Bernstein's theorem \cite{bf73} there exists
a Borel probability measure $m$ on $[0,\infty)$ such that
$$\e^{-\sqrt \lambda}=
\int_0^\infty \e^{-s\lambda} dm(s),$$
 and actually
$\d dm(s)=\frac{1}{2\sqrt\pi}\frac{\e^{-1/(4s)}}{s^{3/2}}ds$. Hence
$$\e^{-t\omega}=\int_0^\infty \e^{-st^2\omega^2}dm(s)=
\frac{1}{2\sqrt\pi}
\int_0^\infty
\frac{t \e^{-t^2/(4s)}}
{ s^{3/2}}
\e^{-s \omega^2} ds.$$
It follows that
$$W(\rx,\ry,|t|)=
\half
\int_{|t|}^\infty dr
(\rho_\rx| \e^{-r\omega} \rho_\ry)
=
\frac{1}{4\sqrt \pi}
\int_{|t|}^\infty dr
\int_0^\infty
\frac{r \e^{-r^2/(4p)}}
{p^{3/2}}
(\rho_\rx| \e^{-ph}\rho_\ry)
 dp.$$
This implies (1) since $\e^{-ph}$ is positivity preserving.

To prove (2), we note that by Proposition \ref{lowbound}
 there exist constants $c_j$ such that
\eq{ko3}
c_1 \e^{c_2t\Delta}(x,y)
\leq \e^{-th}(x,y)
\leq
c_3 \e^{c_4t\Delta}(x,y).
\en
Since $\rho_\rx$ and $\rho_\ry$ are non-negative, we see that by change of variables that ,
$$c_1c_2 W_\infty(\rx,\ry, \sqrt{c_2}|t|)\leq W(\rx,\ry,|t|)\leq c_3c_4 W_\infty(\rx,\ry, \sqrt{c_4}|t|),$$
which completes the proof of the lemma.
\qed

\medskip

Let $\mu_T$ be the probability measure on
$\BR\times \mX$
being absolutely continuous with respect to
$\rP$ such that
\eq{ka7}
d\mu_T=\frac{1}{Z_T}\e^{(q^2/2) \WTT}d\rP,\en
where $Z_T$ denotes the normalizing constant
such that $\mu_T$ becomes a probability measure.
\bl{upper}
One has
\eq{18}
\gamma(T)\leq \EE_{\mu_T}
\left[
\e^{-q^2\WT}
\right].
\en
\el
\proof
Using Theorem \ref{irto} (2) and the shift invariance of $X_{t}$ (see Proposition \ref{main1}) it follows that the denominator of $\gamma(T)$ equals
\[
 \|\e^{- TH}\one\|^{2}=(\one |\e^{- 2T H}\one)= \ix {
\e^{(q^2/2) \int_{-T}^T dt \int_{-T}^T ds
W( X_t, X_s, |t-s|)}}=Z_{T}.
\]
The numerator of $\gamma(T)$ can be estimated by
the Cauchy-Schwarz inequality and shift invariance of $X_{t}$:
\[
\begin{array}{rl}
(\one | \e^{- t H}\one)^{2}=& \left(\ix {
\e^{(q^2/2) \int_{0}^T dt \int_{0}^T ds
W}}\right)^{2}\\[2mm]
\leq & \int \mup \left({\EE}^\rx
\left[\e^{(q^2/2) \int_{0}^T dt \int_{0}^T ds
W}
\right]\right)\left({\EE}^\rx
\left[\e^{(q^2/2) \int_{-T}^0 dt \int_{-T}^0 ds
W}
\right]\right)\\[2mm]
= &\ix {
\e^{(q^2/2) \left(\int_{0}^T dt \int_{0}^T ds
W+\int_{-T}^{0}dt\int_{-T}^{0}dsW\right)}},
\end{array}
\]
where in the last line we use the fact that $X_{s}$ and $X_{t}$ are independent for $s\leq 0\leq t$.
Next we note that if $F(s, t)= F(t,s)$ we have
\[
\begin{array}{rl}
&\int_{0}^{T}ds \int_{0}^{T}d t F(s,t)+ \int_{-T}^{0}ds \int_{-T}^{0}d t F(s,t)\\[2mm]
&=\int_{-T}^{T}ds \int_{-T}^{T}dt F(s, t)- 2\int_{-T}^{0} ds\int_{0}^{T}dt F(s,t).
\end{array}
\]
We can apply this identity to $F(s, t)= W(X_{s}, X_{t}, |t-s|)$ and obtain
\[
\begin{array}{rl}
&\left(\ix {
\e^{(q^2/2) \int_{0}^T dt \int_{0}^T ds
W}}\right)^{2}\\[2mm]
&\leq
\int\mup
{\EE}^{\rx}
\left[\e^{-q^2\int_{-T}^{0}ds\int_{0}^{T} dt W+ (q^2/2)\int_{-T}^{T} ds\int_{-T}^{T}dt W}\right],
\end{array}
\]
which using the definition of $\mu_{T}$ completes the proof of the lemma. \qed

\medskip

Let us take $\lambda$ such that
\eq{assalp}
\frac{1}{\delta +1}<\lambda<1,
\en
where $\delta$ is the exponent in Assumption (E5) and set
\eq{ko11}
A_T:=
\lkk(\rx, \omega)\in \BR\times\mX \
| \
\sup_{|s|\leq T}|X_s(\omega)|\leq T^\lambda\rkk.
\en
The proof of Theorem \ref{hiroshima4} will follow immediately from the following two lemmas.
\bl{yui1}
One has
\[
\lim_{T\rightarrow \infty}
\EE_{\mu_T}\left[
\one_{A_T}\e^{-q^2\WT}\right]=0.
\]
\el
\bl{yui2}
One has
\[
\lim_{T\rightarrow \infty}
\EE_{\mu_T}\left[
\one_{A_T^c}
\e^{- q^2\WT}\right]=0.
\]
\el

{\bf Proof of Theorem \ref{hiroshima4}.}
By Lemmas \ref{yui1}, \ref{yui2} and \ref{upper} it follows that \\ $\lim_{T\to +\infty}\gamma(T)=0$.
We apply then Lemma \ref{auxili}. \qed

\subsection{Proofs of Lemmas \ref{yui1}
and \ref{yui2}}
We prove in this section Lemmas \ref{yui1}
and \ref{yui2}.

{\bf Proof of Lemma \ref{yui1}.}
By Lemma \ref{suzuki}, it suffices to prove that
\beq\label{alti}
 \lim_{T\rightarrow \infty}
\EE_{\mu_T}\left[
\one_{A_T}\e^{- \WTZ}\right]=0.
\eeq
The proof is similar to \cite{lms}.
Let \eqn
&&
\Delta_T=\{(s,t)|0\leq s\leq T, 0\leq t\leq T, 0\leq s+t\leq T/\sqrt 2\},\\
&&
\Delta_T'=\{(s,t)|0\leq s\leq T/\sqrt 2, -s\leq t\leq s\},
\enn
so that
\begin{equation}
\label{arlo}
\begin{array}{rl}
&\int_{-T}^0 dt \int_0^T dt \frac{1}{a^2+|t-s|^2}\\[2mm]
&\geq \int\int_{\Delta_T}ds dt \frac{1}{a^2+|s+t|^2}\\[2mm]
&=
\int\int_{\Delta'_T}dsdt\frac{1}{a^2+s^2}
=\log\lk
\frac{a^2+T^2/2}{a^2}\rk.
\end{array}
\end{equation}
We note now that
$|x-y+ \rx-\ry|^2+|t-s|^2\leq
8T^{2\lambda}+2|x-y|^2+|t-s|^2$
uniformly for $|\rx|\leq T^{\lambda}$, $|\ry|\leq T^\lambda$.
Using (\ref{ne.e6}) and (\ref{arlo})
this yields
\[
\begin{array}{rl}
&\one_{A_T}
\int_{-T}^0 ds \int_0^T dt
W_\infty(X_s,X_t,C_2|s-t|)\\[2mm]
&\geq
\frac{1}{4\pi^2}
\one_{A_T}
\int_{-T}^0 ds \int_0^T dt\int dx dy
\frac{\rho(x)\rho(y)}{
8T^{2\lambda}+2|x-y|^2+C_2|t-s|^2}\\[2mm]
&\geq
\frac{1}{4C_2\pi^2}
\one_{A_T}
\int dxdy\rho (x)\rho (y)
\log\lk
\frac
{8T^{2\lambda}+2|x-y|^2+C_2T^2/2}
{8T^{2\lambda}+2|x-y|^2}
\rk.
\end{array}
\]
Note that $\rho\geq0$ and $\lambda<1$.
Since the right-hand side above goes to $+\infty$ as $T\to \infty$, \kak{alti}
follows. \qed

\medskip

{\bf Proof of Lemma \ref{yui2}.}

Using again Lemma \ref{suzuki} it suffices to prove
\eq{yui4}
\lim_{T\rightarrow \infty}
\EE_{\mu_T}\left[
\one_{A_T^c}
\e^{- \WTZ}\right]=0.
\en
By a change of variables we see that
\[
\int_{-T}^T\int_{-T}^T
\e^{-|s-t|\lambda} ds dt\leq \int_{-\sqrt{2}T}^{\sqrt{2}T}
\int_{-\sqrt{2}T}^{\sqrt{2}T}
\e^{- |t|\lambda} ds dt\leq CT\lambda^{-1}, \ \forall \lambda>0.
\]
Using (\ref{ne.e6b}) and Lemma \ref{suzuki} this implies that
\begin{eqnarray}
&&\label{ch1}
0\leq \int_{-T}^{T}ds \int_{-T}^{T} dt W_{\infty}(X_{s}, X_{t}, C_{2}|s-t|)\leq C \frac{T}{2}\|\hat\rho/|k|\|^{2},\\
&&\label{ch2}
0\leq \int_{-T}^{T}ds \int_{-T}^{T} dt W(X_{s}, X_{t},C_{2}|s-t|)
\leq C \frac{T}{2}\|\hat\rho/|k|\|^{2}.
\end{eqnarray}
Set $\12 \|\hat \rho/|k|\|^2=\xi$.
Hence \kak{ch1}, \kak{ch2} and  the Cauchy-Schwartz inequality yield that
\begin{align*}
&\EE_{\mu_T}
\left[
\one_{A_T^c}
 \e^{-\int_{-T}^{0}ds \int_{0}^{T} dt W_{\infty}}
\right]
\leq
\e^{ TC\xi}
\EE_{\mu_T}
\left[
\one_{A_T^c}
\right]\\
&=
\e^{ TC\xi}
\frac{
\int {\one_{A_T^c}\e^{(q^2/2)\wtt}}d\rP}
{\int{\e^{(q^2/2)\wtt}}d\rP}
\\
&\leq
\e^{ TC\xi}
\frac{
\lk
\int {\e^{ q^2\wtt}}d\rP\rk^\han}
{\int{\e^{(q^2/2)\wtt}}d\rP}
\lk
\int {\one_{A_T^c}} d\rP\rk^\han
\leq
\e^{ TC'\xi}
\lk
\int {\one_{A_T^c}} d\rP\rk^\han.
\end{align*}
By Lemma \ref{lms2} below we know that there exist constants $a,b>0$ such that
\eq{dir}
\int {\one_{A_T^c}} d\rP
\leq T^{-\lambda}(a+bT)^{\12}
\e^{- T^{\lambda(\delta+1)}}.
\en
 Since $\lambda(\delta+1)>1$ this completes the proof of the lemma. \qed

\medskip

\bl{lms2}
There exist constants $a, b>0$ such that \kak{dir} is satisfied,
where $\delta>0$ is the exponent appearing in Assumption (E5).
\el

\subsection{Proof of Lemma \ref{lms2}}
This section is devoted to the proof of Lemma \ref{lms2}.
Let $G\subset \rr^{3}$ be a closed set, and  $T>0$ and $n\in \nn$ are fixed.
We define the stopping time
\begin{equation}
\label{stop}
\tau:= \inf\{T_{j}\ | \ j=0,1,...,n, X_{T_{j}}\in G\}, \quad  T_{j}= \frac{j}{n}T.
\end{equation}
\bl{d7}
Let $\psi\in \HP$ with $\psi\geq 0$ and
$\psi\geq 1$ on $G$.
Let $\tau$ be  in \kak{stop}. 
Then for all $0<\varrho<1$
  one has
$$
\int \mup (\EE^\rx[\varrho^\tau])^2
\leq
(\psi | \psi)+\frac{\varrho^{T/n}}{1-\varrho^{T/n}}
(\psi | (\one-\e^{- (T/n) \lp })\psi).$$
\el
\proof
Set $\psi_{\varrho}(\rx)=\EE^\rx[\varrho^\tau]$.
By the definition of $\tau$ we can see that
\eq{d3}
\psi_{\varrho}(\rx)=1,\quad \rx\in G,
\en
 since $\tau=0$ in the case $X_s$ starts from the inside of $G$.
We can directly see that
\[
 \e^{- s \lp }\psi_{\varrho}(\rx)=\EE^\rx[\EE^{X_s}
[\varrho^\tau]]=
\EE^\rx[\varrho^
{\tau\circ \theta_{s}}]
\]
 by the Markov property,
where $\theta_s$ is the shift on $\mX $ defined by $(\theta_s\omega)(t)=\omega(t+s)$ for $\omega\in \mX $. 
Note that
$(\tau\circ \theta_{T/n})(\omega)=\tau(\omega)-T/n
\geq 0$,
when $\rx=X_0(\omega)\not \in G$.
Hence
\eq{d4}
\varrho^{T/n}\e^{- (T/n) \lp } \psi_{\varrho}(\rx)=\psi_{\varrho}(\rx),\quad
\rx\in G^c.
\en
Clearly
$$
\int \mup (\EE^{\rx}[\varrho^\tau])^2
=(\psi_{\varrho}| \psi_{\varrho})\leq
(\psi_{\varrho}| \psi_{\varrho})+\frac{\varrho^{T/n}}{1-\varrho^{T/n}}(\psi_{\varrho}| (\one-\e^{- (T/n) \lp })\psi_{\varrho}).$$
By \kak{d4} the right-hand side above equals
\eq{d10}
( \one_{G}\psi_{\varrho}| \one_{G}\psi_{\varrho})+\frac{\varrho^{T/n}}{1-\varrho^{T/n}}
( \one_{G}\psi_{\varrho}| (\one-\e^{- (T/n) \lp })\psi_{\varrho}).
\en
Next
\eqn
(\one_{G}\psi_{\varrho}| (\one - \e^{- (T/n) \lp })\psi_{\varrho})&=&
(\one_{G}\psi_{\varrho}| (\one - \e^{- (T/n) \lp })\one_{G}\psi_{\varrho})+ (\one_{G}\psi_{\varrho}| (\one - \e^{- (T/n) \lp })\one_{G^{c}}\psi_{\varrho})\\
&=&(\one_{G}\psi_{\varrho}| (\one - \e^{- (T/n) \lp })\one_{G}\psi_{\varrho})- (\one_{G}\psi_{\varrho}| \e^{- (T/n) \lp }\one_{G^{c}}\psi_{\varrho})\\
&\leq& (\one_{G}\psi_{\varrho}| (\one - \e^{- (T/n) \lp })\one_{G}\psi_{\varrho}),
\enn
since $\e^{- s\lp }$ has a positive kernel.
Hence
\eq{d12}
\int \mup (\EE^{\rx}[\varrho^\tau])^2
\leq
(\psi_{\varrho}\one_G| \psi_{\varrho}\one_G)
+
\frac{\varrho^{T/n}}{1-\varrho^{T/n}}
(\psi_{\varrho}\one_G| (\one-\e^{- (T/n) \lp })\psi_{\varrho}\one_G).
\en
Note that $\psi_{\varrho}(\rx)
\one_G(\rx)\leq \psi(\rx)$ for all
$\rx\in\BR$. Then
\begin{eqnarray}
&&\label{d13}
(\psi_{\varrho}\one_G| \psi_{\varrho}\one_G)
+
\frac{\varrho^{T/n}}{1-\varrho^{T/n}}
(\psi_{\varrho}\one_G| (\one-\e^{- (T/n) \lp })\psi_{\varrho}\one_G)\\
&&
\leq
(\psi| \psi)+
\frac{\varrho^{T/n}}{1-\varrho^{T/n}}
(\psi| (\one-\e^{- (T/n) \lp })\psi).
\non \end{eqnarray}
Then combining \kak{d12} and \kak{d13}
we prove the lemma.
\qed

\medskip

\bp{kv}
Let $\La>0$
and $f\in C(\BR)\cap D(\lp^\han)$.
Then
\eq{d1}
\rP\lk
\sup_{0\leq s\leq T}|f(X_s)|
\geq \La
\rk
\leq \frac{\e}{\La}
\sqrt{(f| f)+T(\lp^\han f| \lp^\han f)}.
\en
\ep
\proof
The proof is a modification of
 \cite[Lemma 1.4 and Theorem 1.12]{var}.
We fix $T>0$ and $n\in \nn$ and define the stopping time $\tau$ as in (\ref{stop}) for the closed set $G:=\{x\in \rr^{3}\ | \ |f(x)|\geq \Lambda\}$.
It follows that
$$\rP\lk\sup_{j=0,...,n}
|f(X_{T_{j}})|\geq \La\rk
=\rP(\tau\leq T).$$
Let $0<\varrho<1$ which will be chosen later. Clearly
\eq{d9}
\rP(\tau\leq T)
\leq
\int {\varrho^{\tau-T}}d\rP
\leq
\varrho^{-T}\int {\varrho^{\tau}}d\rP
\leq
\varrho^{-T} \lk
\int \mup (\EE^\rx[\varrho^\tau])^2\rk^\han.
\en
Let $ \psi\in \HP$ be any function such that $\psi\geq 0$ and
$\psi(x)\geq 1$ on $G$.
Then
applying Lemma~\ref{d7}
  we get \eq{d5}
\int \mup (\EE^\rx[\varrho^\tau])^2
\leq
(\psi| \psi)+\frac{\varrho^{T/n}}
{1-\varrho^{T/n}}(\psi| (\one-
\e^{-(T/n)\lp })\psi).
\en
Since $|f(x)|\geq \Lambda$ on $G$ we can put $\psi= |f(x)|/\La$ in (\ref{d5}) and get
\eq{d88}
\int \mup (\EE^{\rx}[\varrho^\tau])^2
\leq
\frac{1}{\La^2}
(f|f)+\frac{\varrho^{T/n}}{1-\varrho^{T/n}}\frac{1}{\La^2}
(|f|| (\one-\e^{-(T/n)\lp })|f|).
\en
Since
$(|f|| (\one-\e^{-(T/n)\lp })|f|).
\leq
(f| (\one-\e^{-(T/n)\lp })f)$, we have
\eq{d8}
\int \mup (\EE^{\rx}[\varrho^\tau])^2
\leq
\frac{1}{\La^2}
(f|f)+\frac{\varrho^{T/n}}{1-\varrho^{T/n}}\frac{1}{\La^2}
(f| (\one-\e^{-(T/n)\lp })f).
\en
Then by \kak{d9},
$$
\rP\lk\sup_{j=0,...,n}
|f(X_{T_{j}})|\geq \La\rk
\leq
\frac{\varrho^{-T}}{\La}
\left((f|f)+
\frac{\varrho^{T/n}}{1-\varrho^{T/n}}(f| (\one-\e^{-(T/n)\lp })f)
\right)^{\12}.$$
Set $\varrho=\e^{-1/T}$. Then since $\frac{\varrho^{T/n}}{1-\varrho^{T/n}}
\leq n$
\eq{d6}
\rP\lk\sup_{j=0,...,n}
|f(X_{T_j})|\geq \La\rk
\leq
\frac{\e}{\La}
\left((f|f)+
n
(f| (\one-\e^{-(T/n)\lp })f)
\right)^{\12}
\en
follows. Since
$(f| (\one-\e^{-(T/n)\lp })f)\leq (T/n)(\lp^\han f | \lp^\han f)$, we finally get
\eq{d666}
\rP\lk\sup_{j=0,...,n}
|f(X_{T_j})|\geq \La\rk
\leq
\frac{\e}{\La}
\sqrt{(f|f)+
T
(\lp^\han f| \lp^\han f)
}
\en
follows.
We take the limit $n\to \infty$
in the left hand side of \kak{d666}.
By the Lebesgue dominated convergence theorem,
$$\lim_{n\to\infty}
\rP\lk\sup_{j=0,...,n}
|f(X_{T_j})|\geq \La\rk
=
\rP\lk\lim_{n\to\infty}\sup_{j=0,...,n}
|f(X_{T_j})|\geq \La\rk.$$
Since
$f(X_t)$ is continuous in $t$,
$\lim_{n\to\infty}\sup_{j=0,...,n}
|f(X_{T_j})|=\sup_{0\leq s\leq T}|f(X_s)|$ follows.
This completes the proof of the proposition.
\qed

\medskip

{\bf Proof of Lemma \ref{lms2}. }

Let
$f \in C^\infty(\BR)$
such that
$$f (x)=\lkk
\begin{array}{ll}
|x|,& |x|\geq T^\lambda,\\
\leq T^\lambda,&T^\lambda-1<|x|<T^\lambda,\\
0,& |x|\leq T^\lambda-1.
\end{array}\right.
$$
Since $\{x\ |\ f(x)\geq T^\lambda\}= \{x\ | \ |x|\geq T^\lambda\}$ ]
we see that
\eq{va123}
\int {\one_{A_T^c}}d\rP=
\rP\lk
\sup_{|s|\leq T}|X_s|>T^\lambda\rk
=\rP\lk
{\sup_{|s|\leq T}|f(X_s)|>T^\lambda}\rk.
\en
By Proposition \ref{kv}
we have
\eq{va1}
\rP\lk
 {\sup_{|s|\leq T}|f(X_s)|>T^\lambda}\rk
\leq
 \frac{2\e}{T^\lambda}\sqrt{(f,f)+T (\lp^\han f| \lp^\han f)}.
\en
We have
\[
\begin{array}{rl}
(\lp ^{\12}f| \lp ^{\12}f)= &
\q_{0}(f\grp,  f\grp)+ (f\grp| V f\grp)\\[2mm]
\leq &C (\nabla f\grp| \nabla f\grp)+ (f\grp| V f\grp)\\[2mm]
\leq
&C'\|f \nabla \grp\|^{2}+
C''\|\nabla f \cdot \grp\|^{2}+ \|V^{\12}f\grp\|^{2}.
\end{array}
\]
Using the fact that $\supp f\subset \{|x|\geq T^{\lambda}-1\}$, $\nabla f\in O(T^{\lambda})$ and Lemma \ref{ex2}, we obtain
\[
(f| f)+ (L^{\12}f| L^{\12}f)\leq C\e^{- \delta T^{\lambda(\delta+1)}}.
\]
This completes the proof of the lemma.
 \qed

\appendix
\section{Proof of Proposition \ref{main1}}
\label{a4}
In order to prove
Proposition \ref{main1}
we need several steps.
Let $\mB(\BR)$ denotes the
Borel $\sigma$-field.
For $0\leq t_0\leq \cdots\leq t_n$
let the set function
$\nu_{t_0,...,t_n}:\prod_{j=0}^n
\mB(\BR)\to \rr $
be given by
\eq{uchii}
\nu_{t_0,...,t_n}
\lk
\prod_{i=0}^n A_i
\rk
=
(\one_{A_0}| \e^{-(t_1-t_0)\lp }\one_{A_1}
\cdots \e^{-(t_n-t_{n-1}) \lp }\one_{A_n})
\en
and
for $0\leq t$, $\nu_t:\mB(\BR)\to \rr $ by
\eq{uchi}
\nu_t\lk
A\rk=
(\one| \e^{-t\lp}\one_A)
=(\one| \one_A).
\en

{\bf (Step 1)}
The family of set functions
$\{\nu_\xi\}_{\xi\subset\rr ,\#\xi<\infty}$
given by \kak{uchii} and \kak{uchi} satisfies
the consistency condition:
$$
\nu_{t_0,...,t_{n+m}}\lk
\prod_{i=0}^n A_i \times \prod_{i=n+1}^{n+m} \BR
\rk
=\nu_{t_0,...,t_n}\lk
\prod_{i=0}^n A_i\rk
$$ and
by the Kolmogorov extension theorem \cite[Theorem 2.2]{ks}
there exists
a probability measure $\nu_\infty$
on
$((\BR)^{[0, \infty)},\mB((\BR)^{[0,\infty)}))$ such that
\begin{eqnarray}
&&
\label{na3}
\nu_t\lk
A\rk
=
\EE_{\nu_\infty}\left[
 \one_{A}(Y_t)\right],\\
&&
\nu_{t_0,...,t_n}\lk
\prod_{i=0}^n A_i\rk
=
\EE_{\nu_\infty}\left[
\prod_{j=0}^n \one_{A_j}(Y_{t_j})\right], \quad n\geq 1,
\end{eqnarray}
where
$\mB((\BR)^{[0,\infty)})$
denotes the $\sigma$-field generated by cylinder sets,
and $
Y_t(\omega)=\omega(t)$,
$
\omega\in
(\BR)^{[0, \infty)}$,
 is
the coordinate mapping process.
Then
the process $Y=\pro Y$ on
the probability space
$((\BR)^{[0, \infty)}, \mB((\BR)^{[0,\infty)}), \nu_\infty)$
satisfies that
\begin{eqnarray}
&&
\label{t1}
(f_0| \e^{-(t_1-t_0)\lp }f_1
\cdots \e^{-(t_n-t_{n-1}) \lp }f_n)
=\EE_{\nu_\infty}\lkkk
\prod_{j=0}^n f_j(Y_{t_j})\rkkk,\\
&&
\label{t2}
(\one| f)=
(\one| \e^{-t\lp} f)
=
\EE_{\nu_\infty}\lkkk
f(Y_t)\rkkk
=
\EE_{\nu_\infty}\lkkk
f(Y_0)\rkkk
\end{eqnarray}
for $f_j\in L^\infty(\BR)$, $j=0,1,...,n$.

{\bf(Step 2)}
We now see that the process $Y$ has a continuous version.
\bl{n14}
The process $Y$
on
$((\BR)^{[0, \infty)}, \mB((\BR)^{[0,\infty)}), \nu_\infty)$ has a continuous version.
\el
\proof
We note that by \kak{t1},
 \kak{t2}
and
Proposition \ref{6},
$E_{\nu_\infty}[|Y_t-Y_s|^{2n}]$ can be expressed
in terms of the
diffusion process $X^{\rx}=\pro{X^{\rx}}$,
being the solution of
the stochastic differential equation:
\eq{ka6}
X_t^{\rx,j}-X_s^{\rx,j}=
\int_s^t b_j(X_r^{\rx}) dr+
\sum_{k=1}^3
\int _s^t \sigma_{j,k}(X_r^{\rx})\cdot dB_r^k.\quad j=1,2,3.
\en
Since
$$E_{\nu_\infty}[|Y_t-Y_s|^{2n}]=
\sum_{j=1}^3
\sum_{k=0}^{2n}
\lkkk\!
\begin{array}{c} 2n\\ k
\end{array}\rkkk
(-1)^k\EE_{\nu_\infty}\lkkk (Y_t^j)
^{2n-k} (Y_s^j)^k\rkkk,$$
the left hand side above can be express in terms of $\e^{-t\lp}$ as
\begin{align}
&E_{\nu_\infty}[|Y_t-Y_s|^{2n}]\non \\
&=
\sum_{j=1}^3
\sum_{k=0}^{2n}\lkkk\!
\begin{array}{c} 2n\\ k
\end{array}
\!\rkkk (-1)^k
\lk
(x^j)^{2n-k}
\grp|
\e^{-(t-s){K }}
(x^j)^k\grp \rk_{L^2}
\e^{(t-s)\is(\lp)}.
\non \end{align}
Furthermore
by Feynman-Kac formula, i.e., Proposition \ref{6},
the right-hand side above can be expressed
in terms of $X^{\rx}=\pro{X^{\rx}}$ as
\[
\begin{array}{rl}
&\EE_{\nu_\infty}
 [|Y_t-Y_s|^{2n}]\\[2mm]
& =
\int \mup
\Ew
{
 {
 |X_{t-s}^{\rx}-X_0^{\rx}|^{2n}
 \grp(X_0^{\rx})
 \grp(X_{t-s}^{\rx})
 \e^{-\int_0^{t-s}V (X_r^{\rx})dr}
 }} \e^{(t-s)\is(\lp)}.
 \end{array}
 \]
Since $V\geq 0$,
$$
\EE_{\nu_\infty}
 [|Y_t-Y_s|^{2n}]\leq
 \|\grp\|_\infty^2
\e^{(t-s)\is(\lp)}\int \mup
\Ew
{
 {
 |X_{t-s}^{\rx}-X_0^{\rx}|^{2n}
 }} .
$$
We next estimate
$\Ew
{|X_t^{\rx}-X_s^{\rx}|^{2n}}$.
Since $X_t^{\rx,j}$ is the solution to
the stochastic
differential equation \kak{ka6},
we have
$$\Ew{|X_t^{\rx,j}-X_s^{\rx,j}|^{2n}}
\leq 2^{2n-1}
\Ew{\frac{|t-s|^{2n}}{2^{2n}}
\|b_j\|_\infty^{2n}
+\sum_{k=1}^3 \left|
\int_s^t \sigma_{jk}
(X_r^{\rx})dB_r^k \right|^{2n}
}.$$
By the Burkholder-Davies-Gundy inequality \cite[Theorem 3.28]{ks}, we have
\eqn
\Ew{\left|\int_s^t \sigma_{jk}(X_r^{\rx})dB_r^k \right|^{2n}}
\leq
 (n(2n-1))^n |t-s|^n
\|\sigma_{jk}\|_\infty^{2n}.
\enn
Then
$\Ew{
|X_t^{\rx}-X_s^{\rx}|^{2n}}
\leq C |t-s|^n$ with some constant $C$ independent of $s$ and $t$, and
\eq{20}
\EE_{\nu_\infty}\lkkk
|Y_t-Y_s|^{2n}
\rkkk
\leq C |t-s|^n
\en
follows.
Thus $Y=\pro Y$
has a continuous version by Kolmogorov-\v{C}entov continuity theorem \cite[Theorem 2.8]{ks}.
\qed

\medskip

Let
$\ov Y=\pro {\ov Y }$ be the continuous version of $Y$ on
$((\BR)^{[0, \infty)}, \mB((\BR)^{[0,\infty)}), \nu_\infty)$.
The image measure of $\nu_\infty$
on
$(\YYY , \mB(\YYY))$ with
respect to $\ov Y $
is denoted by
$\rQ$, i.e.,
$\rQ=\nu_\infty\circ \ov Y\f$,
and
$\wY _t(\omega)=\omega(t)$
for $\omega\in \YYY $ is the coordinate mapping process.
Then
we constructed a stochastic process $\wY=\pro{\wY} $ on $(\YYY ,
\mB(\YYY), \rQ)$
such that
$\bar Y\stackrel{\rm d}{=}\wY$.
Then
\kak{t1} and \kak{t2}
 can be expressed in terms of $\wY $ as
 \eqn
&&
(f_0| \e^{-(t_1-t_0)\lp }f_1
\cdots \e^{-(t_n-t_{n-1}) \lp }f_n)
=\EE_\rQ\lkkk
\prod_{j=0}^n f_j(\wY _{t_j})\rkkk,\\
&&
(\one| f)=
(\one| \e^{-t\lp} f)
=
\EE_\rQ\lkkk
f(\wY _t)\rkkk
=
\EE_{\rQ}\lkkk
f(\wY _0)\rkkk.
\enn

{\bf(Step 3)}
Define
the regular conditional probability measure on $\YYY $ by
\eq{n15}
\rQ^{\rx}(\cdot)=\rQ(\cdot|\wY _0=\rx)
\en
for
each $\rx\in\BR$.
It is well defined,
since $\YYY $ is a Polish space (completely separable metrizable space). See e.g., \cite[Theorems 3.18. and 3.19]{ks}.
Since the
 distribution of $\wY_0$ equals to $\mup $,
 note
 that
$\rQ(A)
=\int \mup \EE_{\rQ^{\rx}}[\one_A]$.
Then the stochastic
process $\tilde Y=\pro{\tilde Y}$
on $(\YYY , \mB(\YYY), \rQ^{\rx})$
satisfies
\begin{eqnarray}
&&
\label
{ptk}
(f_0| \e^{-(t_1-t_0)\lp }f_1
\cdots \e^{-(t_n-t_{n-1}) \lp }f_n)
=\int \mup \EE_{\rQ^{\rx}}
\left[\prod_{j=0}^n f_j(\wY _{t_j})\right],\\
&&
(\one| \e^{-t\lp}f)
=(\one| f)
=\int dx\grp^2(x) \EE_{\rQ^{\rx}}\lkkk f(\wY_0)\rkkk
=\int \mup f(x).
\end{eqnarray}
\bl
 {I2}
 $\wY$ is a Markov process on $(\YYY ,
 \mB(\YYY), \rQ^{\rx})$
 with respect to the natural filtration
 $\ms M_s=\sigma(\wY_r,0\leq r\leq s)$.
\el
\proof
Let \eq{ptk2}
p_t(\rx, A)=\lk \e^{-t\lp }\one_A\rk(\rx),
\quad A\in \mB(\BR),\quad t\geq0.
\en
Notice that
$p_t(\rx,A)=\Ew{\one_A(X_t^{\rx})}$.
Then the finite dimensional
distribution of $\tilde Y$
is
\eq{finite}
\EE_{\rQ^{\rx}}
\left[\prod_{j=1}^n \one_{A_j}(\wY _{t_j})\right]
=\int \prod_{j=1}^n \one_{A_j}(\rx_j)
\prod_{j=1}^n p_{t_j-t_{j-1}}
(\rx_{j-1}, d\rx_j)
\en
with $t_0=0$ and $\rx_0=\rx$ by
\kak{ptk}.
We show that
$p_t(\rx,A)$ is a probability transition kernel, i.e.,
(1)
$p_t(\rx, \cdot)$ is a probability measure on $\mB(\BR)$,
(2) $p_t(\cdot,A)$ is Borel measurable with respect to $\rx$,
(3) the Chapman-Kolmogorov equality
\eq{chap}
\int p_s(\ry,A)p_t(\rx,d\ry)=p_{s+t}(\rx,A)
\en
is satisfied.
Note that
$\e^{-t\lp }$
is positivity improving.
Then $0\leq \e^{-t\lp } f\leq \one$
for all
function $f$ such that
$0\leq f\leq \one$,
and $\e^{-t\lp }\one=\one$ follows.
Then
 $p_t(\rx,\cdot)$ is the probability measure on $\BR$ with
 $p_t(\rx, \BR)=1$, and (1) follows.
(2) is trivial.
From the semi-group property $\e^{-s\lp}\e^{-t\lp}\one_{A}=
\e^{-(s+t)\lp}\one_A$,
the Chapman-Kolmogorov equality \kak{chap} follows.
Hence $p_t(\rx,A)$ is a probability transition kernel.
We write $\EE$ for $\EE_{\rQ^{\rx}}$ for notational simplicity.
From the identity
$\EE[\one_A(\wY_t)
\EE[f(\wY_r)|\sigma(\wY_t)]]=
\EE[\one_A(\wY_t)f(\wY_r)]$ for $r>t$,
 it follows that
$$\int \one_A(\ry)\EE[f(\wY_r)|\wY_t=\ry]
P_t(d\ry)=\int P_t(d\ry)
\one_A(\ry)
\int
f(\ry')p_{r-t}(\ry, d\ry'),$$
where $P_t(d\ry)$ denotes the distribution
of $\wY_t$ on $\BR$. Thus
$$\EE[f(\wY_r)|\wY_t=\ry]
=\int f(\ry')p_{r-t}(\ry,d\ry')$$
follows a.e. $\ry$
with respect to $P_t(d\ry)$.
Then
$\EE[f(\wY_r)|\sigma(\wY_t)]=
\int f(\ry) p_{r-t}(\wY_t,d\ry)$
and
\eq{t3}
\EE[\one_A(\wY_r)|\sigma(\wY_t)]=
p_{r-t}(\wY_t,A)
\en
follow.
By using \kak{t3}, \kak{finite} and the Chapman-Kolmogorov equality \kak{chap},
we can show that
$$
\EE\lkkk \one_{A}(\wY_{t+s})
\prod_{j=0}^n \one_{A_j}(\wY_{t_j})\rkkk
=
\EE\lkkk \EE\lkkk
\one_{A}(\wY_{t})|\sigma(\wY_s)\rkkk \prod_{j=0}^n \one_{A_j}(\wY_{t_j})\rkkk$$
for $t_0\leq\cdots\leq t_n\leq s$. This implies that
$
\EE[\one_A(\wY_{t+s})|\ms M_s]=
\EE[\one_A(\wY_{t})|\sigma(\wY_s)]$.
Then $\wY $ is Markov with respect to the natural filtration under the measure $\rQ^{\rx}$.
\qed

\medskip

{\bf(Step 4)}
We extend $\wY =\pro{\wY }$
to a process on the whole real line $\rr $.
Set
$\tilde\mX_+=\YYY \times \YYY $,
$\tilde{\ms M}=\mB(\YYY)\times\mB(\YYY)$ and
${\tilde \rQ}^{\rx}=\rQ^{\rx}\times \rQ^{\rx}$.
Let $(\tilde X_t)_{t\in\rr }$
be the stochastic process on
the product space
$(\tilde\mX_+, \tilde {\ms M}, \tilde \rQ^{\rx})$,
defined by
for $\omega=(\omega_1,\omega_2)\in
\tilde\mX_+$,
\eq{I3}
\tilde X_t(\omega)=\lkk
\begin{array}{ll}
\wY_t(\omega_1),& t\geq 0,\\
\wY_{-t}(\omega_2),& t<0.
\end{array}
\right.
\en
Note that
 $\tilde X_0=\rx$ almost surely with respect to
 $\tilde \rQ^{\rx}$ and
$\tilde X_t$ is continuous in $t$ almost surely.
It is trivial to see that
$\tilde X_t$, $t\geq0$,
and $\tilde X_s$, $s\leq 0$,
 are independent,
 and $\tilde X_t\stackrel{\rm d}{=}
 \tilde X_{-t}$.

{\bf(Step 5)}
{\bf Proof of Theorem \ref{main1}:}\\
The image measure of $\tilde \rQ^{\rx}$ on
$(\mX , \mB(\mX))$ with respect to $\tilde X$ is denoted by
${\rP^{\rx}}$, i.e.,
\eq{ka9}
{\rP^{\rx}}=\tilde \rQ^{\rx}\circ\tilde X\f.
\en
Let
$
X_t(\omega)=\omega(t)$,
$t\in\rr $,
$\omega\in{\mX }$, be the coordinate
mapping process.
Then we can see that
\eq{cent}
X_t\stackrel{\rm d}{=}\tilde Y_t\quad
(t\geq 0),
\quad \quad
X_t\stackrel{\rm d}{=}\tilde Y_{-t}\quad
(t\leq 0).
\en
Since by (Step 3), $\pro {\tilde Y}$ and
$(\tilde Y_{-t})_{t\leq 0}$ are
Markov processes with respect to
the
natural filtration
$\sigma(\tilde Y_s, 0\leq s\leq t)$ and
$\sigma(\tilde Y_s, -t\leq s\leq 0)$, respectively,
 $\pro X$
 and
$(X_t)_{t\leq 0}$
are also Markov processes with respect to
$(\FFF _t^+)_{t\geq 0}$ and
$(\FFF _t^-)_{t\leq 0}$,
respectively.
Thus the Markov property (4) follows.
We also see that
$(X_{s})_{s\leq 0}$
and $\pro X$ are
independent and
$X_{-t}\stackrel{\rm d}{=}
X_t$
by \kak{cent} and (Step 4).
Thus reflection symmetry (3) follows.
\bl{final}
Let $-\infty <t_0\leq t_1\leq \cdots\leq
t_n$. Then
\eq{I4}
\ixp {f_0(X_{t_0})\cdots f_n(X_{t_n})}
=(f_0, \e^{-(t_1-t_0)\lp}f_1\cdots \e^{-(t_n-t_{n-1})\lp} f_n).
\en
\el
\proof
Let $t_0\leq\cdots \leq t_n\leq 0\leq t_{n+1}\leq\cdots t_{n+m}$.
Then we have by the independence of
$(X_{s})_{s\leq 0}$
and $\pro X$,
\[
\begin{array}{rl}
&
\ixp {f_0(X_{t_0})\cdots f_{n+m}(X_{t_{n+m}})}\\[2mm]
&=
\ixp
{f_0(X_{t_0})\cdots f_{n}(X_{t_{n}})}
\EE_{{\rP^{\rx}}}\lkkk
f_{n+1}(X_{t_{n+1}})\cdots f_{n+m}(X_{t_{n+m}})\rkkk.
\end{array}
\]
Since we have
\begin{align}
\label{I5}
&\EE_{{\rP^{\rx}}}\lkkk
f_{n+1}(X_{t_{n+1}})\cdots f_{n+m}(X_{t_{n+m}})\rkkk\\
&=
\lk
\e^{-t_{n+1}\lp}
f_{n+1}
\e^{-(t_{n+2}-t_{n+1})\lp}
f_{n+2}\cdots \e^{-(t_{n+m}-t_{n+m-1})\lp}f_{n+m}\rk(x)\non
\end{align}
and
\begin{align}
\label{I6}
& \EE_{{\rP^{\rx}}}\lkkk
f_{0}(X_{t_0})\cdots f_{n}(X_{t_{n}})\rkkk\\
&=
\EE_{{\rP^{\rx}}}\lkkk
f_{0}(\wY_{-t_0})\cdots f_{n}(\wY_{-t_{n}})\rkkk \non\\
&=
\lk
\e^{+t_{n}\lp}
f_{n}
\e^{-(t_{n}-t_{n-1})\lp}
f_{n-1}\cdots \e^{-(t_{1}-t_{0})\lp}f_1
\rk(x),\non
\end{align}
by \kak{I5} and \kak{I6} we obtain that
\begin{align*}
&
\ixp {f_0(X_{t_0})\cdots f_{n+m}(X_{t_{n+m}})} \\
&=
(\e^{+t_{n}\lp}
f_{n}
\cdots \e^{-(t_{1}-t_{0})\lp}f_1,
\e^{-t_{n+1}\lp}
f_{n+1}
\cdots \e^{-(t_{n+m}-t_{n+m-1})\lp}f_{n+m}) \\
&=
(f_1,
\e^{-(t_1-t_0)\lp}
f_2
\cdots
\e^{-(t_{n+m}-t_{n+m-1})\lp}f_{n+m}).
\end{align*}
Hence \kak{I4} follows.
\qed

\medskip

From Lemma \ref{final}
it follows that
for any $s\in\rr $,
$$\ixp{\prod_{j=0}^ nf_j(X_{t_j})}
=
\ixp{\prod_{j=0}^ nf_j(X_{t_j+s})}.$$
Hence shift invariance (5) is obtained.
\qed

\medskip

\catcode`~=11 
\newcommand{\urltilde}{\kern -.15em\lower .7ex\hbox{~}\kern .04em}
\catcode`~=13 

\noindent {\bf Acknowledgments:}\\
FH acknowledges support of Grant-in-Aid for
Science Research (B) 20340032
from JSPS and
Grant-in-Aid for Challenging Exploratory Research 22654018
from JSPS, and is thankful to the hospitality of Universit\'e de Paris XI, where part of this work has been done.

\end{document}